\documentclass[12pt]{iopart}
\usepackage{graphicx}
\usepackage{color}
\begin{document}

\bibliographystyle{unsrt}

\hyphenation{photo-emission photo-electric Photo-electric photo-electrons photo-electron}

\title[Charging of dust grains by flowing plasmas and photo\-emission]{Charging of insulating and conducting dust grains by flowing plasma and photo\-emission}
\author{W J Miloch$^{1,2}$, S V Vladimirov$^2$, H L P\'{e}cseli$^3$ and J Trulsen$^1$}

\address{$^1$ Institute of Theoretical Astrophysics, University of Oslo, Box 1029 Blindern, N-0315 Oslo, Norway}
\address{$^2$ School of Physics, The University of Sydney, Sydney, NSW 2006, Australia}
\address{$^3$ Department of Physics, University of Oslo, Box 1048 Blindern, N-0316 Oslo, Norway}

\eads{\mailto{w.j.miloch@astro.uio.no}}

\date{\today}

\begin{abstract}
The charging of conducting or alternatively insulating dust grains in a supersonic plasma flow with a directed photon flux is studied by the particle-in-cell method. The electron emission modifies the surface charge distribution on a grain and in the surrounding plasma. The charge and potential distributions on and around a dust grain are studied for different photon fluxes and different angles of the incident unidirectional photons with respect to the plasma flow velocity vector. Continuous and pulsed radiations are considered. We show that photo\-emission allows controlling the charge on conducting grains, and discuss the charging of stationary and spinning insulating grains. Interactions between positively charged grains can be stronger than for negatively charged grains. The simulations are carried out in two spatial dimensions, treating ions and electrons as individual particles.
\end{abstract}
\pacs{52.27.Lw, 52.65.Rr}
\submitto{NJP}
\maketitle

\section{Introduction}
The charging of an object in a plasma is one of the basic problems in plasma physics. The understanding of this process is important in studies of interactions between the plasma and the object, or between many objects in a plasma. The question is particularly important in studies of dusty plasmas, where a number of charged dust grains can form ordered structures, such as dust clusters, strings, and crystals \cite{Shukla_Mamun_2002,Vladimirov_Ostrikov_2005, Ishihara_2007}.
In dusty plasma experiments, grains are usually levitated in the sheath region above the electrode, and they are charged negatively due to the high mobility of electrons. Such grains can be exposed to an ion flow.  A wake and characteristic regions of enhanced ion density (ion focus) are observed behind dust grains immersed in flowing plasmas \cite{Vladimirov_Nambu_1995,Melzer_Schweigert_1996, Ivlev_Morfill_1999, Miloch_Pecseli_Trulsen_2008}. The ion focusing and the corresponding potential enhancement are more conspicuous for supersonic flows \cite{Miloch_Pecseli_Trulsen_2008}, and can lead to the alignment of grains in a direction of the flow \cite{Melzer_schweigert_1999, Maiorov_Vladimirov_2000,Vladimirov_Maiorov_2003a, Hebner_Riley_2004, Samarian_Vladimirov_2005}. Analogous problems can be formulated also for larger objects moving with respect to a plasma, such as spacecrafts or meteoroids \cite{Svenes_Troim_1994,Melandso_Goree_1995}.

In a space environment, dust is often exposed to electromagnetic radiation \cite{Horanyi_1996}. Radiation can be directional or isotropic, either due to background radiation or scattering of directed light \cite{Hayakawa_Yamashita_1969}. The situation is relevant not only for dust in space, but also for dusty surfaces of larger lunar bodies. In the latter case, the dust on the surface is charged by the plasma and directed solar radiation. It has been argued that the shadowing of light can lead to strong electric fields and transport the dust above the lunar surface \cite{Wang_Horanyi_2007}. In laboratory plasmas, the radiation is either due to the plasma glow or an external light source, thus it is either isotropic or directed,  similarly to the space environment \cite{Sickafoose_Colwell_2000}. 
If the energy of incoming photons is larger than the work function of the dust surface material, the photo\-electron current contributes to the net current to the grain  and should be included in the charging analysis \cite{Ishihara_2007, Weingartner_Draine_2001, Klumov_Vladimirov_2005, Klumov_Vladimirov_2007}. 
In several respects the physics of this process resembles that for electron emissive probes \cite{Schrittwieser_Ionita_2008}.
 The differences between the two physical processes are found in the mechanisms for the electron emissions and to some extent also in the velocity distributions of the emitted electrons.
Photo\-emission will change the total charge on the dust grain and the surface charge distributions, and it can lead to new types of interactions between dust grains. Structures comprising positively charged dust grains in a plasma in the presence of UV radiation have been discussed theoretically \cite{Rosenberg_Mendis_1995, Rosenberg_Mendis_1996}, and observed in experiments \cite{Fortov_Nefedov_1998, Samarian_Vaulina_2000}.

A theory describing the charging of dust grains with photo\-emission in a self-consistent way is difficult to develop. In particular, photo\-electrons can modify the plasma in the vicinity of dust grains. Several theoretical studies consider simplified models \cite{Rosenberg_Mendis_1996, Khrapak_Nefedov_1999, Ostrikov_Yu_2001}. To study more realistic problems, one should employ numerical simulations, which can account for non-linear and other possible phenomena, and model the charging of a dust in plasma in a self-consistent way. By numerical simulations, it was demonstrated that the charge of the dust cloud in a plasma discharge can be modified by UV radiation \cite{Land_Goedheer_2007}. The potential structures around a positively biased spacecraft were also studied numerically \cite{Engwall_Eriksson_2006}. In these works the electrons were treated as a Boltzmann distributed background. Neither of these studies considered the self-consistent charging of isolated objects. 

One of our intentions here is to provide a more realistic model by including electrons (photo\-electrons in particular) in the analysis. In a recent communication, we have demonstrated that UV radiation allows for an accurate control of the charge on an isolated conducting grain \cite{Miloch_Vladimirov_2008}. We have also showed that photo\-electrons can modify and polarize the surrounding plasma, enabling stronger interactions between positively charged conducting dust grains. 

In the present paper, we study numerically the charging of conducting or alternatively insulating dust grains in a supersonic plasma flow with a directed photon flux. We analyze the charge, density and potential distributions for different fluxes and energies of photons,  and for different angles between the incoming unidirectional photons and the plasma flow velocity vector. Continuous as well as pulsed radiation is considered. Using a particle-in-cell (PIC) method, we simulate the entire charging process in a collision\-less plasma. The simulations are carried out in two spatial dimensions, treating ions and electrons as individual particles. We consider unidirectional radiation, which is relevant for a dust in space exposed to the solar radiation, as for example lunar dust, or a laboratory experiment with an external radiation source.

\section{Numerical code}
We have modified the numerical particle-in-cell (PIC) code used in our previous studies \cite{Miloch_Pecseli_Trulsen_2008, Miloch_Pecseli_Trulsen_2007, Miloch_Vladimirov_2008b}, by including a photon flux and the photo\-electric effect \cite{Miloch_Vladimirov_2008}. 
We consider collisionless plasmas in a two-dimensional system in Cartesian coordinates. 
Both electrons and ions are treated as individual particles, with the electron to ion temperature ratio $T_e/T_i=100$, and $T_e=0.18~\mathrm{eV}$.  The ion to electron mass ratio is $m_i/m_e=120$ in most of the simulations. As a control case we analyze also results for a conducting grain with $m_i/m_e=36720$ (to represent Neon). The plasma density is $n=10^{10}~\mathrm{m^{-2}}$, and the plasma flow velocity is $v_d=1.5~C_s$, with $C_s$ denoting the speed of sound. Because of the large thermal velocity of electrons, the plasma flow is represented solely by the ion drift.

A circular dust grain of radius of $R=0.375$ in units of the electron Debye length $\lambda_{De}$  is placed inside a simulation box of size of $50 \times 50$ $\lambda_{De}$. The grain is assumed to be massive and immobile, except for the simulations of the spinning insulator. Initially, the grain is charged only by the collection of electrons and ions. For a perfectly insulating grain, a plasma particle hitting the dust grain surface remains at this position at all later times and contributes to the surface charge distribution. To model a small conductor in this work, the charge is redistributed equally on the dust grain surface at each time step. Such an algorithm is simple to use and is also found in other numerical studies \cite{Lapenta_1999}, but it does not account for the electric dipole moment on the conducting dust grain as induced by the anisotropic potential distribution in flowing plasmas. The equally distributed surface charge will not necessarily cancel electric fields inside the grain, and thus the algorithm is not adequate for grains larger than the Debye length or for grains of shapes different than spherical (or circular). 
This algorithm is different form the one used  in our previous studies of the charge distribution on larger dust grains, which enforced constant potential within the dust grain \cite{Miloch_Pecseli_Trulsen_2008, Miloch_Pecseli_Trulsen_2007}. The computational expenses of that algorithm were lengthy simulations and strict constraints on shapes and sizes of simulated dust grains.

A directed photon flux is switched on after approximately 40 ion plasma periods $\tau_i$. At this time, we can assume that the surface charge distribution on a grain has reached a stationary level. The code is run typically up to 50 ion plasma periods.
Three different angles between the incoming photons and the direction of the ion drift are considered: $\alpha=\{ 0^{\circ}, 90^{\circ}, 180^{\circ} \}$. For conducting grains, the simulated photon flux is $\Phi_{h \nu} \in (0.25, 2.5) \times 10^{19}~\mathrm{m^{-2}s^{-1}}$. This together with photon energies $E_{h\nu}$ of $4.8$, $5.5$ and $7.2~\mathrm{eV}$, gives a photon power density $H \in (1.9, 28.8) ~\mathrm{Wm^{-2}}$. The work function $W$ of the conducting dust grains is taken to be $W=4.5~\mathrm{eV}$, which is close to work functions of many metallic materials \cite{Rosenberg_Mendis_1996}. For insulating grains, the photon energies $E_{h\nu}$ are $10.3$, $11.0$ and $12.7~\mathrm{eV}$. This, together with the photon fluxes as for the case of conducting grains, gives a photon power density of $H \in (4.5, 50.8) ~\mathrm{Wm^{-2}}$. The work function of the insulating grain is taken as $W=10~\mathrm{eV}$, which implies that photo\-electrons will have the same energies as for the conducting case.

When a photon hits the surface of the dust grain, a photo\-electron of energy $E=E_{h\nu}-W$ is produced at distance $l=sv\Delta t$ from the dust grain surface, where $s$ is an uniform random number $s \in (0,1]$,  $\Delta t$ is the computational time step and $v$ is the photo\-electron speed. Photo\-electron velocity vectors are uniformly distributed over the hemicircle and directed away from the dust grain surface, that is in accordance with Lambert's law.

To investigate the stability of the surface charge distribution on insulating grains, we simulate also spinning grains. Instantaneous rotation of angles $\beta=\{1^{\circ},5^{\circ},10^{\circ} \}$, as well as continuous rotation with the angular velocities $\Omega = \{\pi/180, \pi/36, \pi/18,  \pi/2, \pi, 2\pi \}$ in units of $\mathrm{rad}/\tau_i$  (corresponding to the grain rotation by angles of $1,5,10,90,180,$ and 360 degrees within $\tau_i$, respectively) are considered. The rotation starts at approximately one ion plasma period after the onset of radiation. As a control case we also rotate the grain throughout the whole simulation.

\section{Numerical results}

The present section is in two parts.  First we consider the charging of a conducting or alternatively insulating grain in the presence of continuous radiation. This problem is followed by the results from the simulations with pulsed radiation.

\subsection{Continuous radiation}
The charge on a conducting dust grain exposed to a continuous photon flux becomes more positive with the onset of radiation and saturates within one ion plasma period. Some of the results for a conducting grain have been presented before \cite{Miloch_Vladimirov_2008}, but we include them also in the present work for completeness. The saturation charge on a conducting grain, which is  summarized in Table \ref{tab:uv_charging_c}, depends on the flux density and photon energy. For a sufficiently high photon flux, the grain becomes positively charged. For low fluxes, the saturation charge does not depend significantly on the photon energy. For higher fluxes, high energy photo\-electrons lead to a more positive dust grain. The relative fluctuations of the charge are largest for the grain with the smallest charge. The absolute and relative charge fluctuations are smallest for the case without photo\-emission.
The results for the total charge for $E_{h\nu}=7.2~\mathrm{eV}$ are very similar to the case of $E_{h\nu}=5.5~\mathrm{eV}$, and therefore they are not presented in Table~\ref{tab:uv_charging_c}.

\Table{The total charge $q_t$ on a conducting dust grain for different photon energies $E_{h\nu}$ and different photon fluxes $\Phi_{h\nu}$ for $\alpha=0^{\circ}$, averaged over $10\tau_i$. The relative charge fluctuations $\Delta q_t$ are also shown. The total charge $q_t$ is normalized with the unitary two-dimensional charge  $q_{0}=e\left[ n_{0(3D)}\right] ^{1/3}$, where $e$ is an elementary charge, and $n_{0(3D)}$ is the plasma density in the corresponding three-dimensional system. The unit of $q_{0}$ is $[q_{0}]=\mathrm{C/m}$.
\label{tab:uv_charging_c}
\lineup}
\br
&\centre{2}{$E_{h\nu}$=4.8~{eV}}&\centre{2}{$E_{h\nu}=5.5~\mathrm{eV}$}\\
&\crule{2} & \crule{2}\\
$\Phi_{h\nu}$ & $q_t $ & $\Delta q_t $ & $q_t $ & $\Delta ~q_t $\\
$(10^{19}\mathrm{m^{-2}s^{-1}})$ & ${(q_0)}$ & $(\%)$ & $ {(q_0)}$ & $(\%)$\\
\mr
0.0 & \-755 & \0\04 & \0\-755 & \0\04 \\
0.25 & \-163 & \019 & \0\-168 & \017\\
0.50 & \019 & 173 & \0\012 & 258 \\
1.25 & 251& \018 & \0273 & \018\\
2.50 & 795 & \0\08 & 1330& \0\07 \\
\br
\end{tabular}
\end{indented}
\end{table}


The floating potential on a positively charged grain for two highest photon fluxes is shown in Table~\ref{tab:flpot} together with the corresponding results from analytical calculations. The analytical results  for the floating potential in Table~\ref{tab:flpot} are calculated for a balance of the photo\-emission $i_{h\nu}$, ion $i_i$, and electron $i_e$ currents to the grain:  $i_e=i_i+i_{h\nu}$. For consistency, we restrict our analysis to a two-dimensional case. The photo\-emission current can for this case be expressed by \cite{Shukla_Mamun_2002}:
\begin{equation}
i_{h\nu}=A_{h\nu}\Phi_{h\nu(2D)}e \exp \left( -\frac{e\Psi}{kT_{h \nu}} \right),
\end{equation}
where $e>0$, and it is assumed that the photo\-electric yield and photo\-emission efficiency equal unity. In the present two-dimensional model with unidirectional photons, $A_{h\nu}=2R$, and $\Phi_{h\nu (2D)}=c (\Phi_{h\nu(3D)}/c)^{2/3}$, with the physical dimension of  $[\Phi_{h\nu(2D)}]=m^{-1}s^{-1}$. Subscripts $(2D)$ and $(3D)$ stand for two-dimensional and three-dimensional cases, respectively. The ion current to a plane surface segment with area $A_i$ due to singly charged ions drifting at supersonic speed $v_d$, can be approximated by
\begin{equation}
i_{i}= A_{i} n_{0(2D)} v_d e \exp \left( -\frac{e\Psi} {kT_{i}} \right),
\end{equation}
where we define the ion cross section for supersonic ion flow as $A_i=2R$. The ion current is consistent with the current to a probe for retarding fields \cite{Schott_1968}, but we replaced the ion thermal velocity by $v_d$ and neglected the numerical constant by assuming that ion velocities  are unidirectional and normal to the probe surface at the sheath edge. We note that the ion current to the positively charged grain is negligible due to small thermal velocity of ions, but nevertheless we include it in the calculations for completeness. Since the grain radius $R$ is comparable to the electron Debye length, we use a general expression for the orbit-motion-limited (OML) current  to the conducting cylinder, to calculate the electron current to the grain \cite{Schott_1968}: 
\begin{eqnarray}
\fl i_{e}=-\frac{1}{4}A_{e} n_{0 (2D)} e v_{the} \frac{r_s}{R} \left[ \mathrm{erf}\left( \sqrt{\frac{-\gamma}{r^2_s/R^2-1}} \right) + \right. \nonumber\\
\left. +\frac{R}{r_s}\exp(-\gamma) \left( 1-  \mathrm{erf}\left( \sqrt{\frac{-\gamma r^2_s}{r^2_s-R^2}}\right) \right)\right],
\end{eqnarray}
where $\gamma=-e\Psi/kT_e$, $A_e=2\pi r$, and $r_s$ is the sheath radius, which in our calculations is set to $r_s=3R$. We introduced the error function as $\mathrm{erf}(x)=2/\sqrt{\pi} \int_{0}^{x} \exp (-y^2) dy$.

\Table{The floating potential on a grain for different photon energies $E_{h\nu}$ and different photon fluxes $\Phi_{h\nu}$ for $\alpha=0^{\circ}$. The results from the simulations $\Psi_{fl,~\mathrm{sim}}$ as well as from analytical calculations  $\Psi_{fl,~\mathrm{calc}}$ are shown. \label{tab:flpot} \lineup}
\br
&\centre{2}{$E_{h\nu}$=4.8~{eV}}&\centre{2}{$E_{h\nu}=5.5~\mathrm{eV}$}\\
&\crule{2} & \crule{2}\\
$\Phi_{h\nu}$ & $\Psi_{fl,~\mathrm{sim}}$ & $\Psi_{fl,~\mathrm{calc}}$ & $\Psi_{fl,~\mathrm{sim}}$ & $\Psi_{fl,~\mathrm{calc}}$   \\
$(10^{19}\mathrm{m^{-2}s^{-1}})$ & $\mathrm{(V)}$ & $\mathrm{(V)}$ & $\mathrm{(V)}$ & $\mathrm{(V)}$ \\
\mr
1.25 & 0.17 & 0.17 & 0.28 & 0.27 \\ 
2.50 &  0.16 & 0.36 & 0.56 & 0.66 \\
\br
\end{tabular}
\end{indented}
\end{table}

\begin{figure}[!htb]
\begin{center}
\includegraphics[width=0.9\columnwidth]{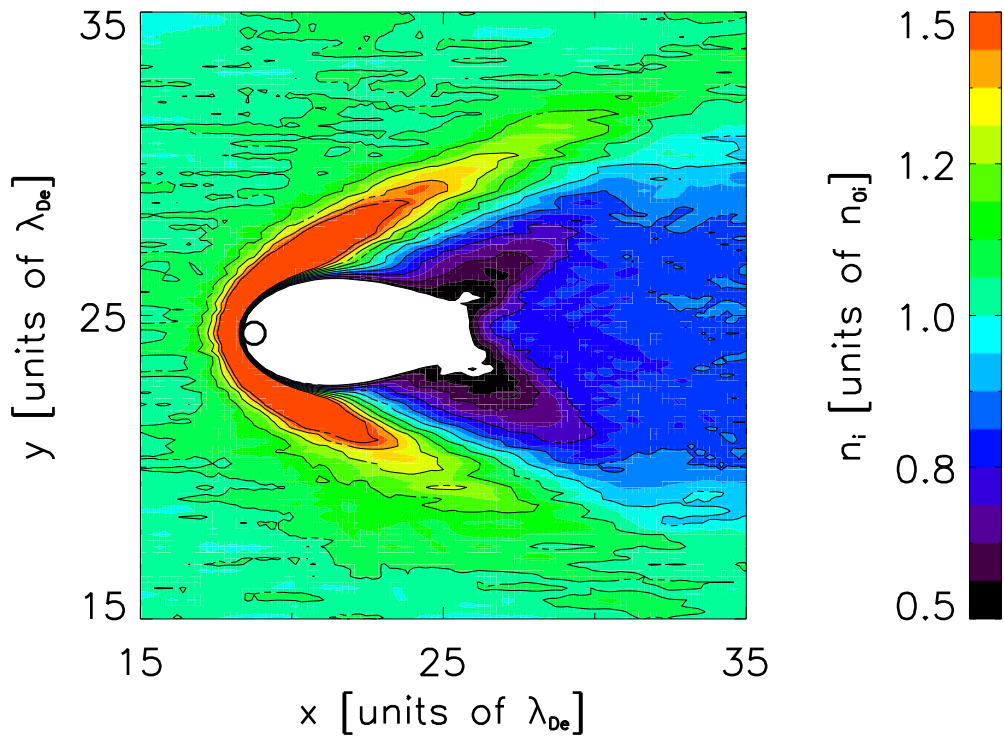}
\caption{The ion density around a conducting dust grain exposed to the photon flux $\Phi_{h\nu} = 2.5 \times 10^{19}~\mathrm{m^{-2}s^{-1}}$ of energy $E_{h\nu}=4.8~\mathrm{eV}$ averaged over nine ion plasma periods $\tau_i$. $\alpha=0^{\circ}$ and the plasma flow is in the positive $x$ direction. The white region corresponds to ion densities below $0.5n_{0i}$.}
\label{fig:uv_ionwake_c}
\end{center}
\end{figure}

The density and potential distributions around a conducting dust grain depend on the flux and energy of the photons. For photon fluxes of $\Phi_{h\nu}=0.25\times 10^{19}~\mathrm{m^{-2}s^{-1}}$, when the grain is negatively charged, we observe an ion focusing in the wake \cite{Miloch_Pecseli_Trulsen_2008}. The ion density in the focusing region is $n_i \approx 1.2n_{0i}$, where $n_{0i}$ is the undisturbed ion density far from the grain. This result is smaller than for the corresponding case without photo\-emission where we had $n_i \approx 2.2n_{0i}$. The ion focusing is destroyed for positively charged grains. In this case, ions are slowed down and deflected in front of the grain.
Consequently, a region of an enhanced ion density is formed in front of the grain, while downstream from the grain there is a distinct boundary between the wake and the undisturbed plasma, see Fig.~\ref{fig:uv_ionwake_c}.
The shape of the enhanced ion density region depends on $\alpha$: it is more pronounced and located closer to the dust grain surface for $\alpha=0^{\circ}$, and further from the grain for $\alpha=180^{\circ}$. For $\alpha=90^{\circ}$, an asymmetry in the enhanced ion density is observed \cite{Miloch_Vladimirov_2008}. 

The wake in the ion density behind a conducting grain (a region where $n_i < 0.5 n_{0i}$), scales with the photon flux and  photon energy, increasing for increasing fluxes and energies. The measured spatial extent of the wake is summarized in Table~\ref{tab:uv_wake_c}. The ion wake corresponds to the white region behind the grain in Fig.~\ref{fig:uv_ionwake_c}.

\Table{The width $w$ and length $d$ of the ion wake behind a positively charged conducting dust  grain for different photon energies $E_{h\nu}$ and different photon fluxes $\Phi_{h\nu}$ for $\alpha=0^{\circ}$. The unit of $w$ and $d$ is the electron Debye length $\lambda_{De}$. The ion wake was not observed for $\Phi_{h\nu} < 0.5 \times 10^{19}~\mathrm{m^{-2}s^{-1}}$. \label{tab:uv_wake_c} \lineup}
\br
&\centre{2}{$E_{h\nu}$=4.8~{eV}}&\centre{2}{$E_{h\nu}=5.5~\mathrm{eV}$}&\centre{2}{$E_{h\nu}=7.2~\mathrm{eV}$}\\
&\crule{2} & \crule{2} & \crule{2}\\
 $\Phi_{h\nu}$ & $w$ & $d$  & $w$ & $d$ & $w$  & $d$ \\
$(10^{19}\mathrm{m^{-2}s^{-1}})$ & $(\lambda_{De})$  & $(\lambda_{De})$ & $(\lambda_{De})$ & $(\lambda_{De})$ & $(\lambda_{De})$  & $(\lambda_{De})$  \\
\mr
0.50 & 0.7 & 3.1 & 0.7 &  \03.5 & 0.9 & \03.7\\
1.25 & 2.1 & 6.8 & 2.3 & \07.5 & 2.5 & \07.3 \\
2.50 & 3.6 & 7.0 & 5.9 & 11.1 & 6.3 & 12.7\\
\br
\end{tabular}
\end{indented}
\end{table}

\begin{figure}[!htb]
\begin{center}
\includegraphics[width=0.9\columnwidth]{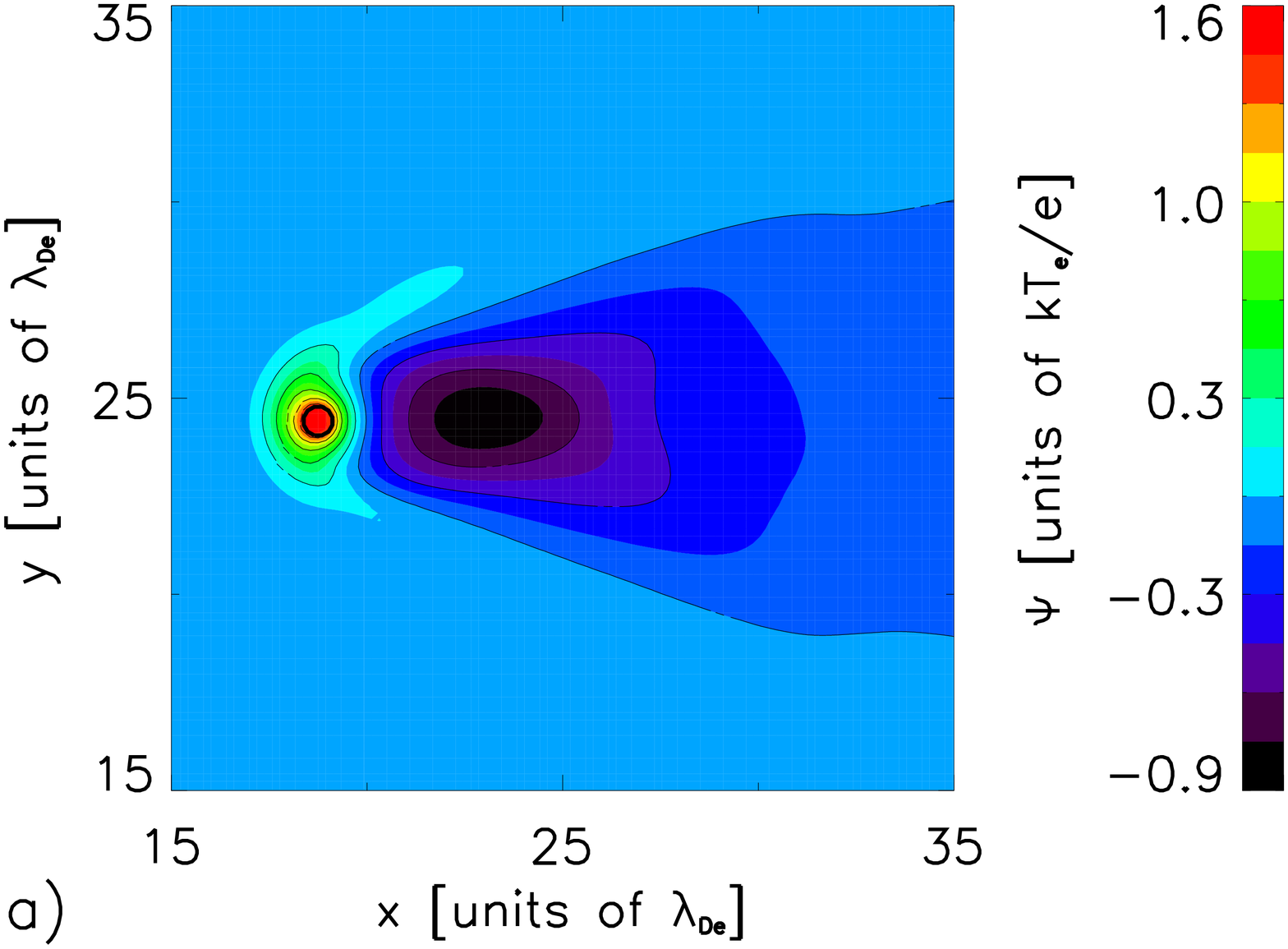}
\includegraphics[width=0.9\columnwidth]{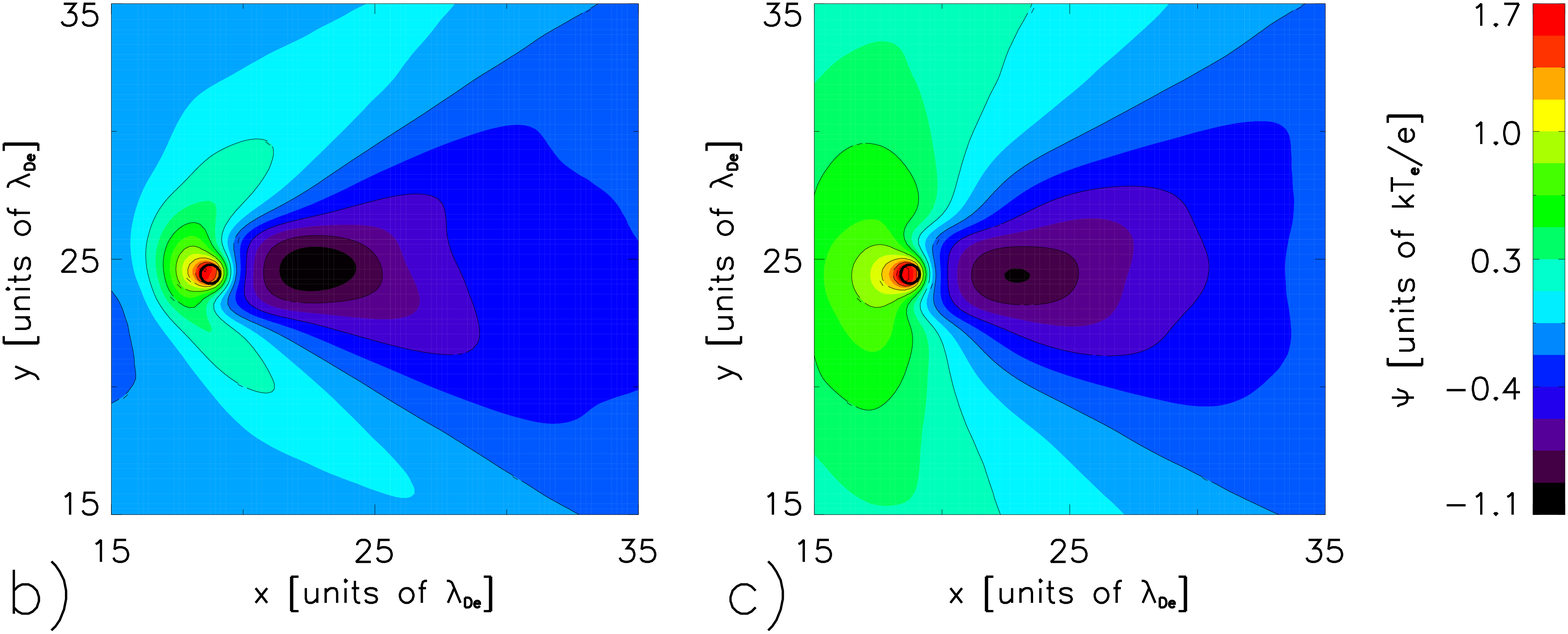}
\caption{The potential around a conducting dust grain exposed to the photon flux $\Phi_{h\nu} = 2.5 \times 10^{19}~\mathrm{m^{-2}s^{-1}}$ of energy $E_{h\nu}=4.8~\mathrm{eV}$ for $\alpha=0^{\circ}$ (a), $\alpha=90^{\circ}$ (b), and $\alpha=180^{\circ}$ (c). The data were averaged over a time interval of nine ion plasma periods $\tau_i$. The plasma flows  in the positive $x$ direction.}
\label{fig:polarization_c}
\end{center}
\end{figure}

The potential around the positively charged conducting dust grain is polarized for higher photon fluxes. In Fig.~\ref{fig:polarization_c}, the potential distribution around the conducting dust grain is shown for different angles of incidence of photons with the flux $\Phi_{h\nu} = 2.5 \times 10^{19}~\mathrm{m^{-2}s^{-1}}$. The potential is negative behind, and positive in front of the grain. The polarization of the plasma is most conspicuous for $\alpha=180^{\circ}$.
\

The results from simulations with a more realistic ion mass are in accordance with the results obtained with the reduced ion mass. For a realistic ion mass, the total charge $q_t$ on a grain without photo\-emission is $q_t=-1883q_0$. This result is more negative than for simulations with reduced ion mass. The ratio of the saturation charges for different ion masses is $q_{t,1}/q_{t,2}=2.5$. It is close to the ratio

\begin{equation}
\frac{Q_{0,1}}{Q_{0,2}}=\frac{ {\ln  \left( \gamma_1/2\pi  +1 \right)} } { \ln \left( \gamma_2/2\pi +1 \right)}=2.9,
\end{equation}
where indices $1,2$ refer to different ion to electron mass ratios for ion masses $m_i=36720m_e$, and $m_i=120m_e$ respectively, and $Q_0$ is a theoretical charge on a grain in a stationary plasma  in a two-dimensional system, given by

\begin{equation}
Q_0= 2 \pi \epsilon_0 \Psi_{fl} \frac{R}{\lambda_D}\frac{K_1(R/\lambda_D)}{K_0(R/\lambda_D)}.
\label{q_theory}
\end{equation} 

In (\ref{q_theory}), $K_0$ and $K_1$ are modified Bessel functions, $R$ is the radius of a grain, $\gamma=m_i/m_e$, and $\Psi_{fl}$ is a floating potential of the grain, here given by

\begin{equation}
\Psi_{fl}=-\frac{\kappa T_e}{2e}\left[ \ln \left( \frac{\gamma}{2\pi} +1 \right) \right].
\label{psi_theory}
\end{equation}

In (\ref{psi_theory}) it is assumed that cold ions are reaching the surface of the large conducting object at the Bohm speed. A more detailed discussion on Equations (\ref{q_theory}) and (\ref{psi_theory}) is given elsewhere \cite{Miloch_Pecseli_Trulsen_2007}.

Without photo\-emission, ions are streaming out of the ion focus with a wider angle for realistic ion masses as compared to the case with a reduced ion mass. This is due to different ion drift velocities for the two cases. In both cases the ion drift is $v_d=1.5~C_s$, with the speed of sound given by $C_s=\sqrt{\kappa (T_e+5T_i/3)/m_i}$, in the plasma far from the grain. In the vicinity of the grain, the plasma parameters are modified due to particle trapping and sheath formation. With photo\-emission, the saturation charge and the wake are similar for both ion masses. The length of the wake is the same, while the width for larger ion masses is larger by $5\%$. The charge saturates within one ion plasma period for both cases. The ion plasma period for the realistic ion mass is approximately 20 times larger than for the reduced ion mass. With another simulations for grains with the radius of $2R$, we find that the saturation charge is approximately twice the charge value for the grain with radius of $R$.

\begin{figure}[!htb]
\begin{center}
\includegraphics[width=0.9\columnwidth]{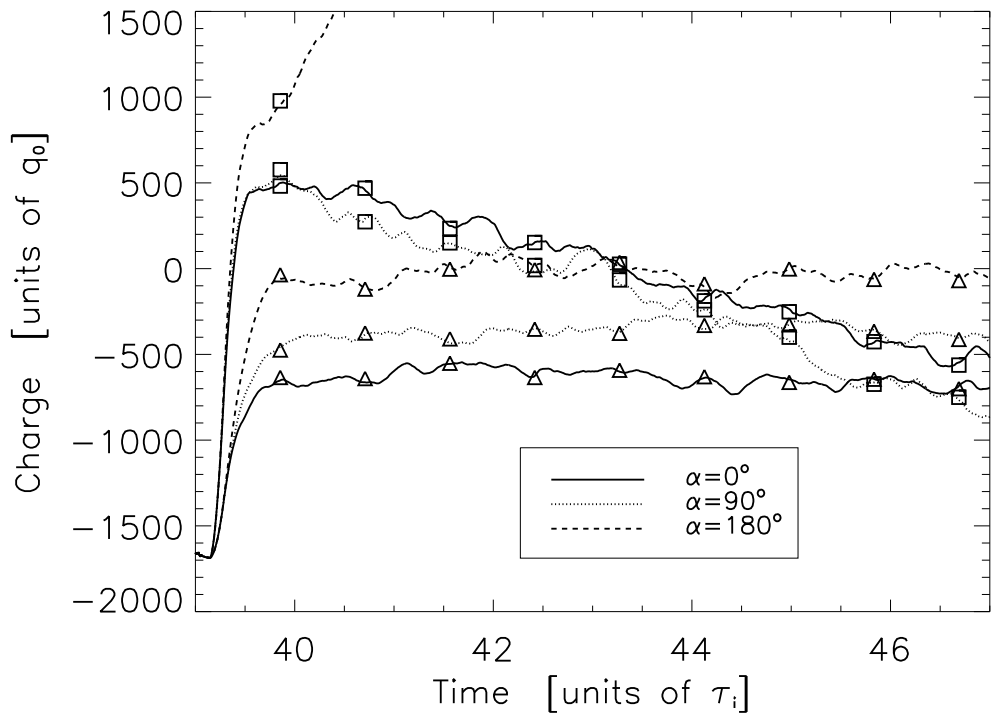}
\caption{The total charge on an insulating dust grain as a function of time for different photon fluxes and angles of photon incidence $\alpha$. Squares correspond to the photon flux  $\Phi_{h\nu} = 2.5 \times 10^{19}~\mathrm{m^{-2}s^{-1}}$, triangles to  $\Phi_{h\nu} = 0.5 \times 10^{19}~\mathrm{m^{-2}s^{-1}}$. The photon energy is $E_{h\nu}=11.0~\mathrm{eV}$. The results are smoothed with a moving box average filter for presentation.}
\label{fig:uv_charging_i}
\end{center}
\end{figure}

The charging of an insulating grain exposed to a continuous radiation differs from the conducting case. The saturation in the charging characteristics is observed for photon fluxes $\Phi_{h\nu}=0.25 \times 10^{19}\mathrm{m^{-2}s^{-1}}$ and $\Phi_{h\nu}=0.50 \times 10^{19}\mathrm{m^{-2}s^{-1}}$, when the total charge on the dust grain remains negative. For photon fluxes and energies high enough to change the sign of the total charge on the grain, the charge does not saturate within the time-span of our simulations. In all cases, the charging depends on the angle of incidence, see Fig.~\ref{fig:uv_charging_i}. For lower fluxes the charge is getting less negative with increasing $\alpha$. For higher fluxes, the charge can become positive, and then negative again within a few ion plasma periods. This is not the case for $\alpha=180^{\circ}$, for which the charge increases towards more positive values.

\begin{figure}[!htb]
\begin{center}
\includegraphics[width=0.9\columnwidth]{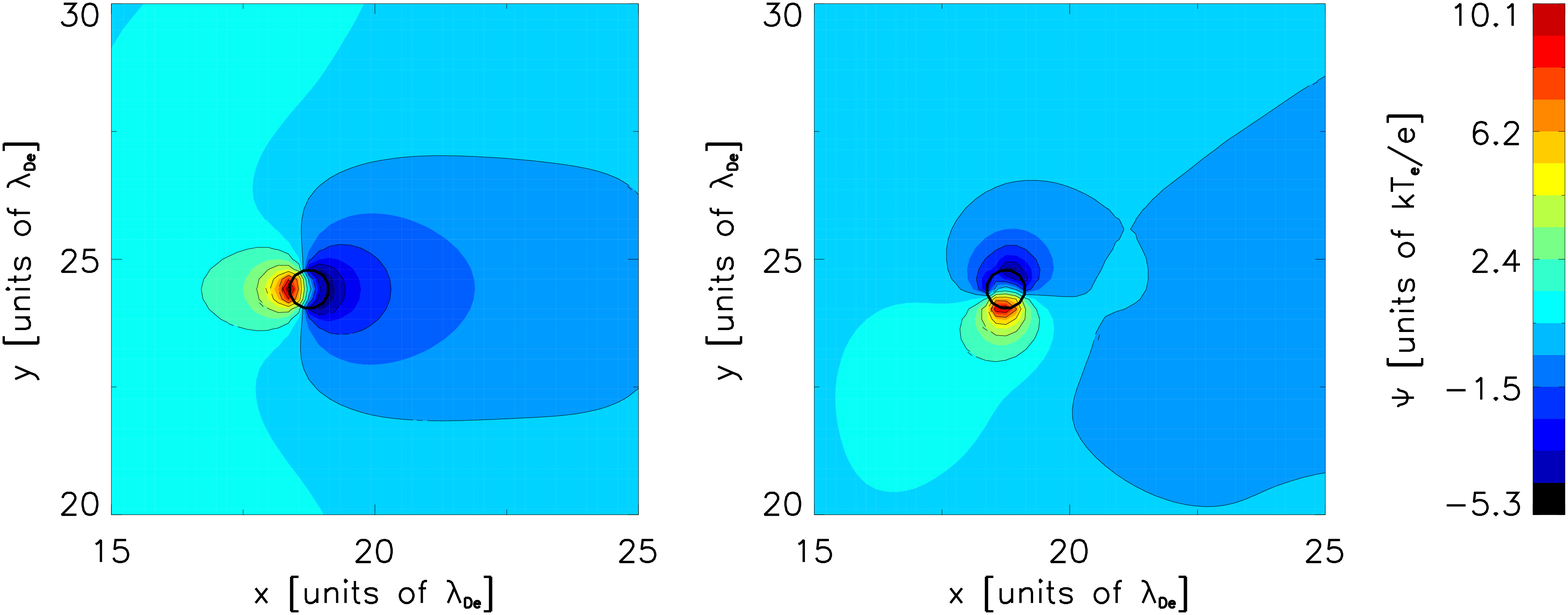}
\includegraphics[width=0.9\columnwidth]{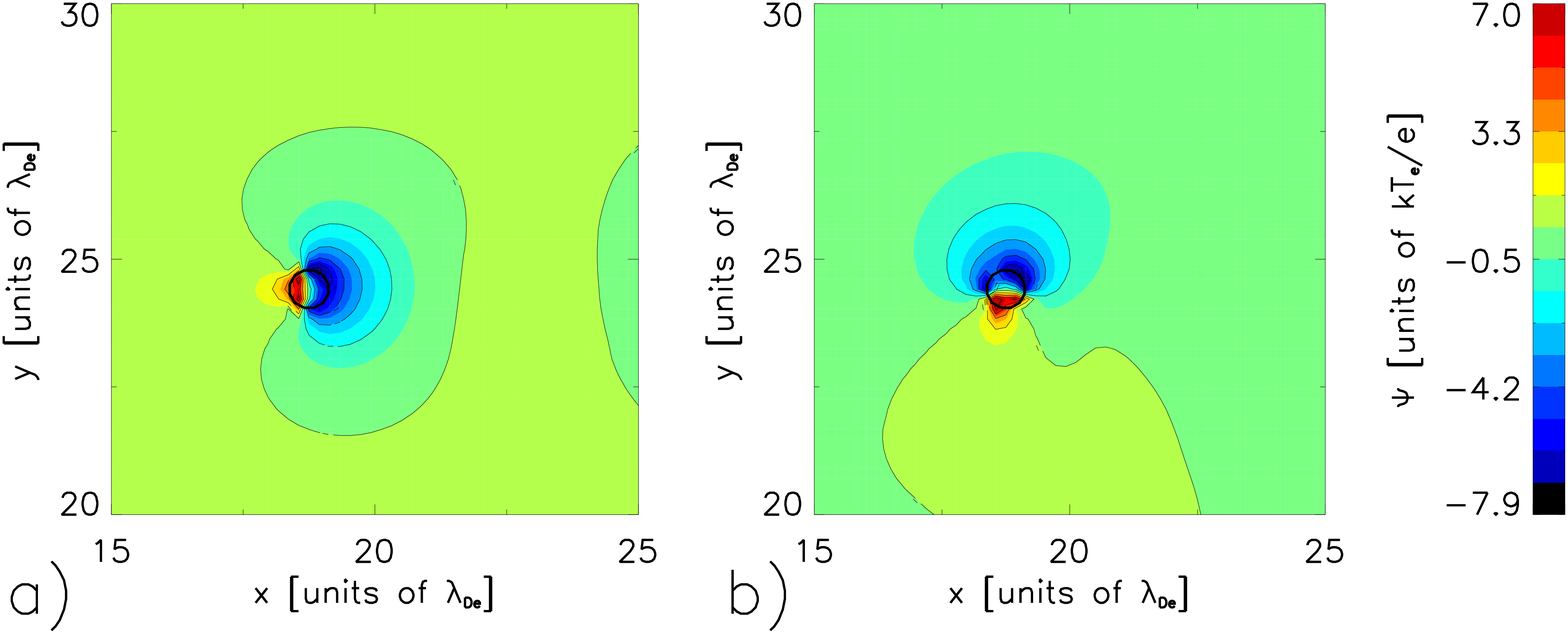}
\caption{
The potential around an insulating dust grain exposed to the photon flux $\Phi_{h\nu} = 2.5 \times 10^{19}~\mathrm{m^{-2}s^{-1}}$, $\alpha=0^{\circ}$ (a) and $\alpha=90^{\circ}$ (b) of energy $E_{h\nu}=11.0~\mathrm{eV}$ averaged over two ion plasma periods: $t \in ( 39.5,41.5 ) \tau_i$ (top) and $t \in ( 48.0,50.0 ) \tau_i$ (bottom). The plasma flows in the positive $x$ direction.}
\label{fig:polarization_i}
\end{center}
\end{figure}

With the onset of the photon flux, we observe the development of an electric dipole moment on the grain  which is antiparallel to the direction of the incident photons, see Fig.~\ref{fig:polarization_i}. This electric dipole moment due to the photo\-electrons does not saturate for high photon fluxes, and it is stronger than the electric dipole moment induced by the ion flow.

\begin{figure}[!htb]
\begin{center}
\includegraphics[width=0.9\columnwidth]{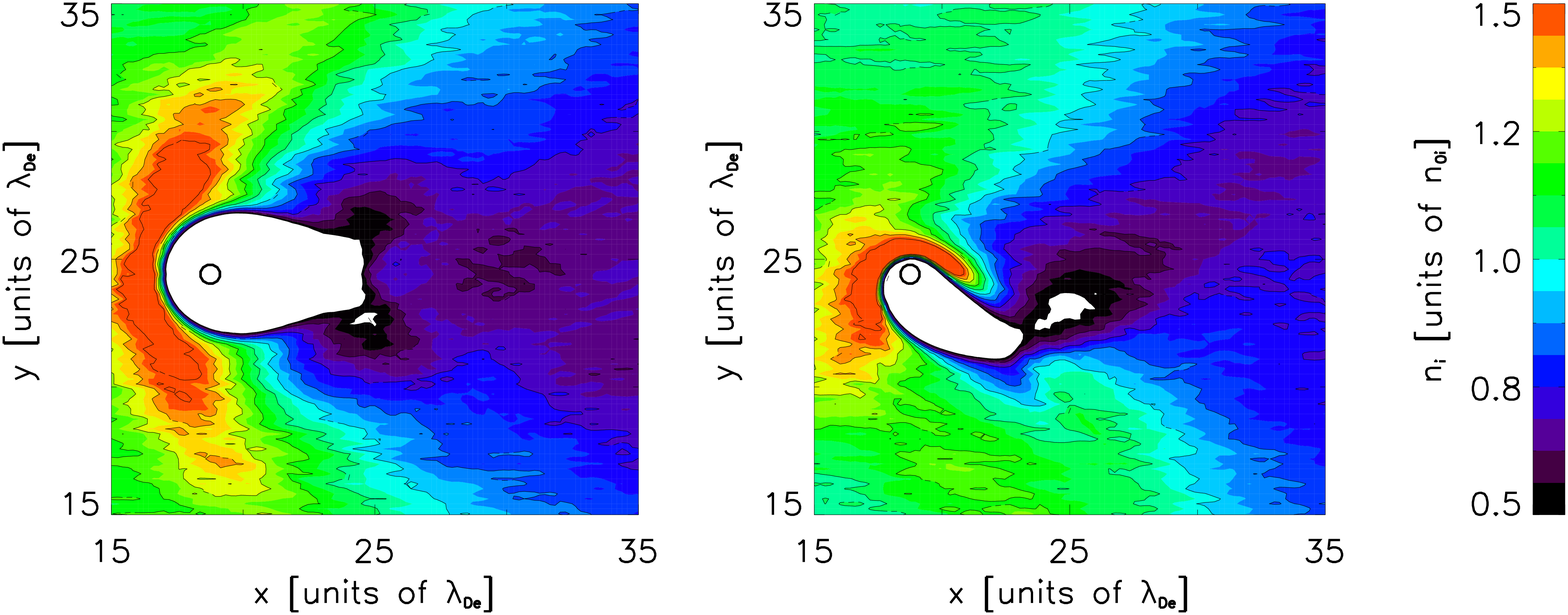}
\includegraphics[width=0.9\columnwidth]{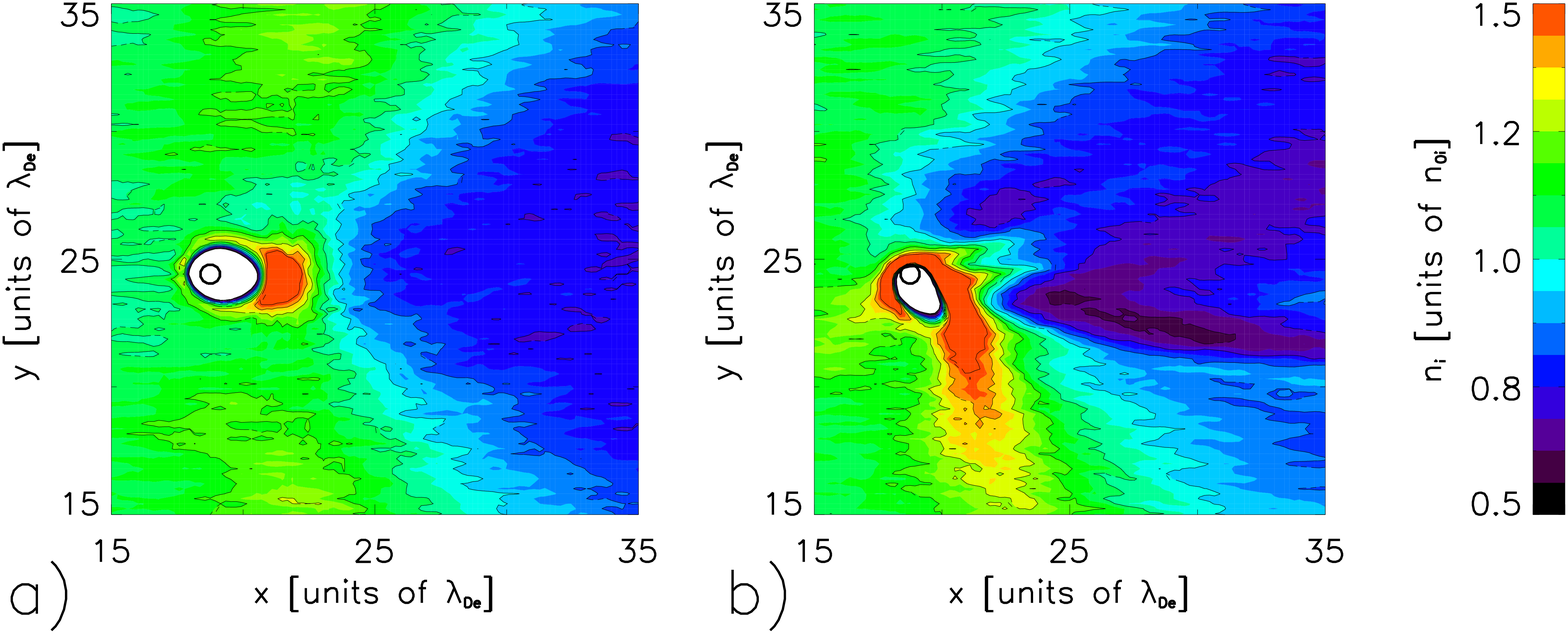}
\caption{The ion density around an insulating dust grain exposed to the photon flux $\Phi_{h\nu} = 2.5 \times 10^{19}~\mathrm{m^{-2}s^{-1}}$, $\alpha=0^{\circ}$ (a) and $\alpha=90^{\circ}$ (b) of energy $E_{h\nu}=11.0~\mathrm{eV}$ averaged over two ion plasma periods: $t \in ( 39.5,41.5 ) \tau_i$ (top) and $t \in ( 48.0,50.0 ) \tau_i$ (bottom). The plasma flows in the positive $x$ direction. White regions correspond to ion density levels below $0.5 n_{0i}$.}
\label{fig:uv_ionwake_i}
\end{center}
\end{figure}

The density and potential distributions around an insulating grain evolve in time, see Figs.~\ref{fig:polarization_i} and \ref{fig:uv_ionwake_i}. The potential distribution is influenced by the electric dipole moment due to photo\-emission. This moment becomes smaller when the total charge is negative. For a positively charged grain, the ion focusing region in the wake is destroyed, and the wakes behind the dust grain with $\alpha=0^{\circ}$ and $\alpha=180^{\circ}$ are similar to the conducting case. The wake is strongly asymmetric for $\alpha=90^{\circ}$. For high photon fluxes, when the charge on the grain reaches negative values, the wake behind insulator becomes smaller, and the ion focus can be retrieved. The asymmetric charge distribution for $\alpha=90^{\circ}$ is present also after the closure of the wake, as shown in Fig.~\ref{fig:uv_ionwake_i}b).

\begin{figure}[!htb]
\begin{center}
\includegraphics[width=0.6\columnwidth]{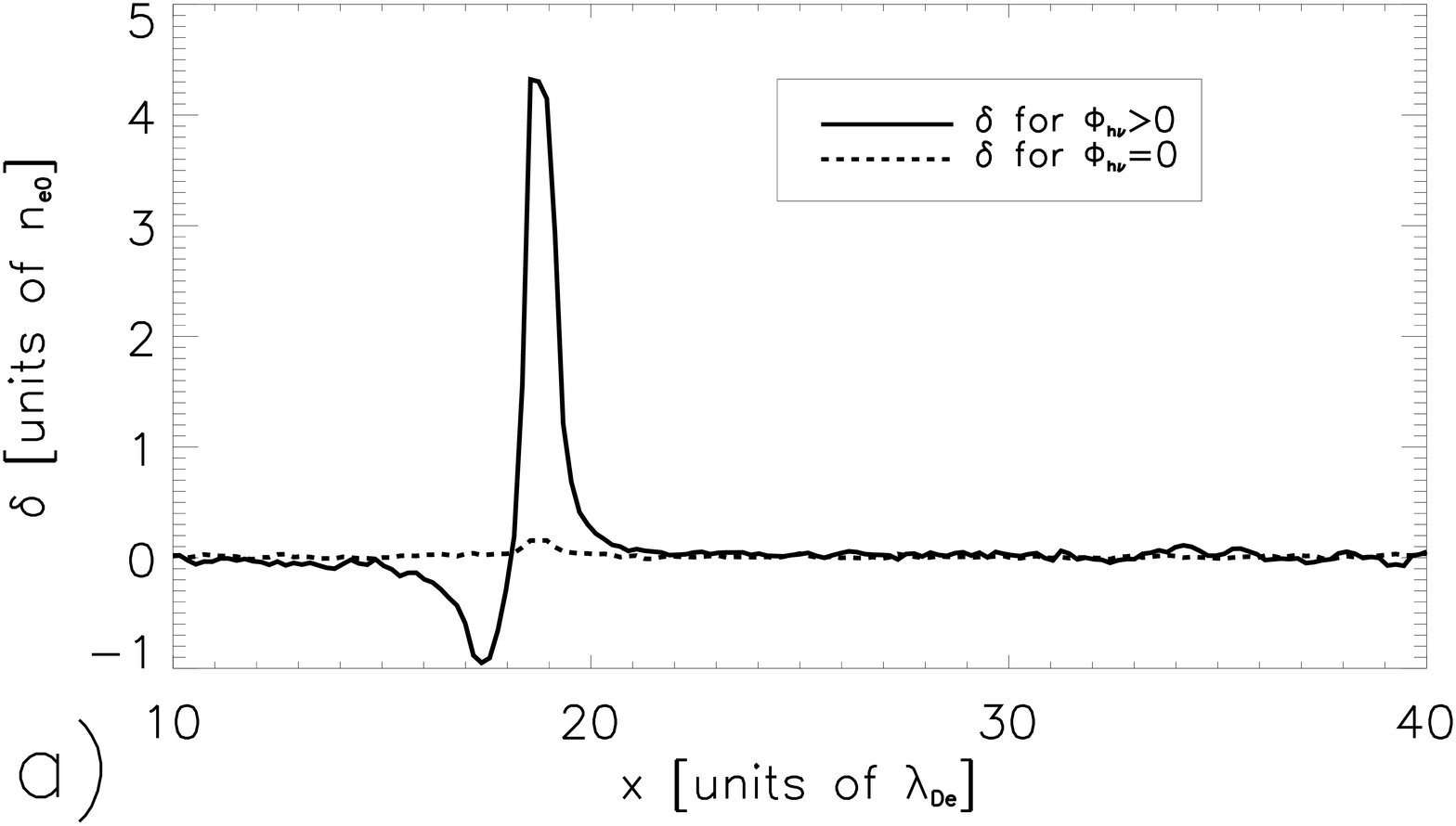}
\includegraphics[width=0.6\columnwidth]{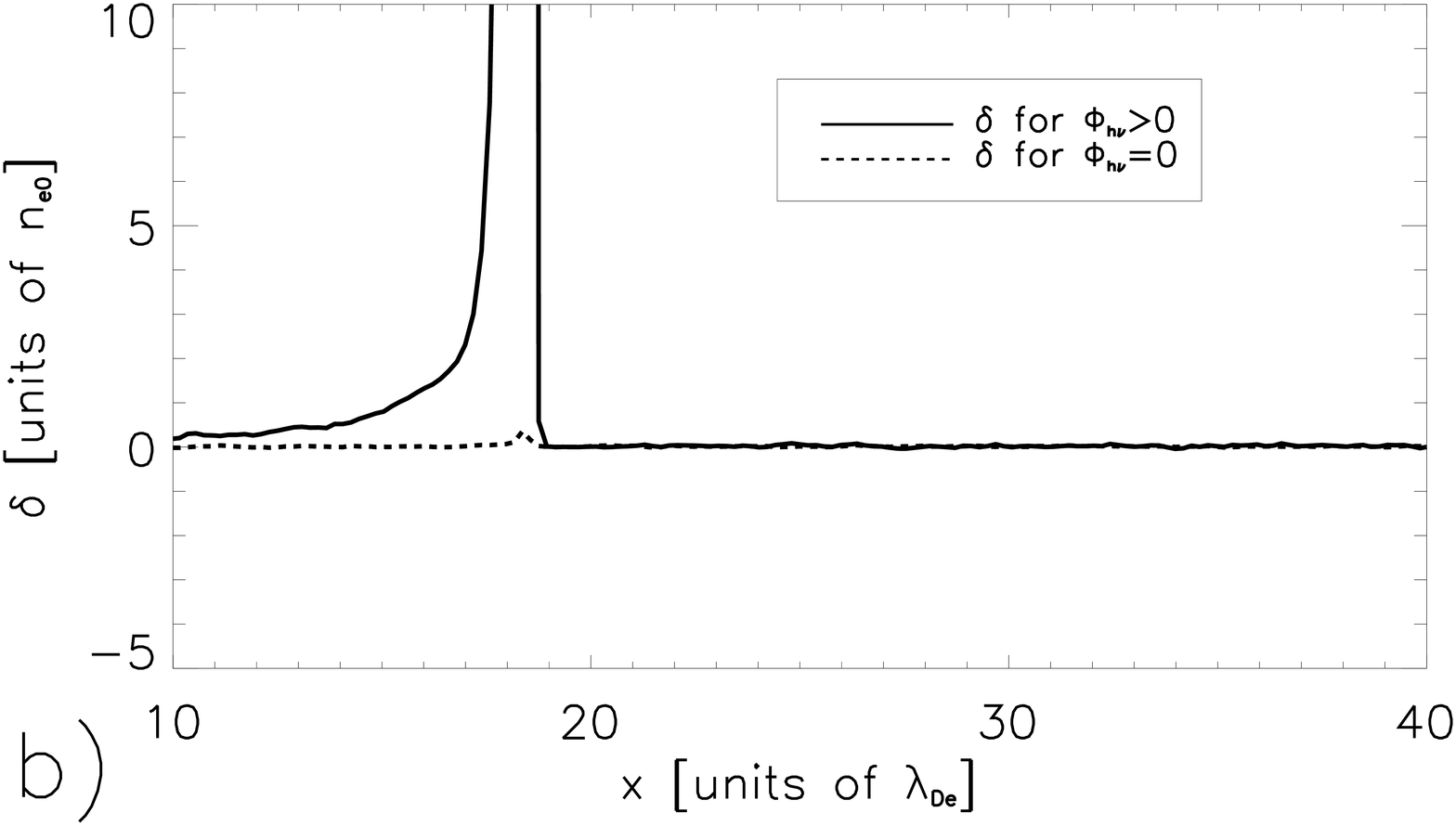}
\caption{The difference $\delta$ between the density of Boltzmann distributed electrons that would correspond to the calculated potential and the actual electron density is shown for the case with (solid line) and without (dashed line) photo\-emission. Both conducting (a) and insulating (b) dust grains are considered for $\alpha=0^{\circ}$.}
\label{fig:uv_Boltzmann}
\end{center}
\end{figure}

In Fig.~\ref{fig:uv_Boltzmann}, we illustrate the difference $\delta$ between the density of Boltzmann distributed electrons that would correspond to the calculated potential and the actual electron density: $\delta=n_{e0}\exp[e\Psi/kT_e]-n_e$, where $e>0$ is the magnitude of the electron charge. Results for both conducting and insulating stationary grains are shown in Fig.~\ref{fig:uv_Boltzmann}. Before the onset of the photon flux the electrons can be well approximated by the Boltzmann distribution. With photo\-emission, the electrons are no longer Boltzmann distributed. The largest discrepancies for conductors are associated with a surplus of electrons due to the photo\-electron emission, and to a region of an enhanced ion density in front of the dust grain, where electrons are underrepresented. For insulators the electric dipole governs the potential in vicinity of the grain.

Instantaneous rotation of the insulating dust grain with an angle $\beta$ has little effect on the grain charging characteristic. For $\alpha=180^{\circ}$, no significant change is observed in the potential and density distributions, while for $\alpha=0^{\circ}$ the rotation leads to weak asymmetries there. The asymmetries are more pronounced for larger $\beta$. For $\alpha=0^{\circ}$ the charging characteristics are similar to the case without rotation, but the charge becomes more negative at a slightly slower rate with increasing $\beta$. 

\begin{figure}[!htb]
\begin{center}
\includegraphics[width=0.59\columnwidth]{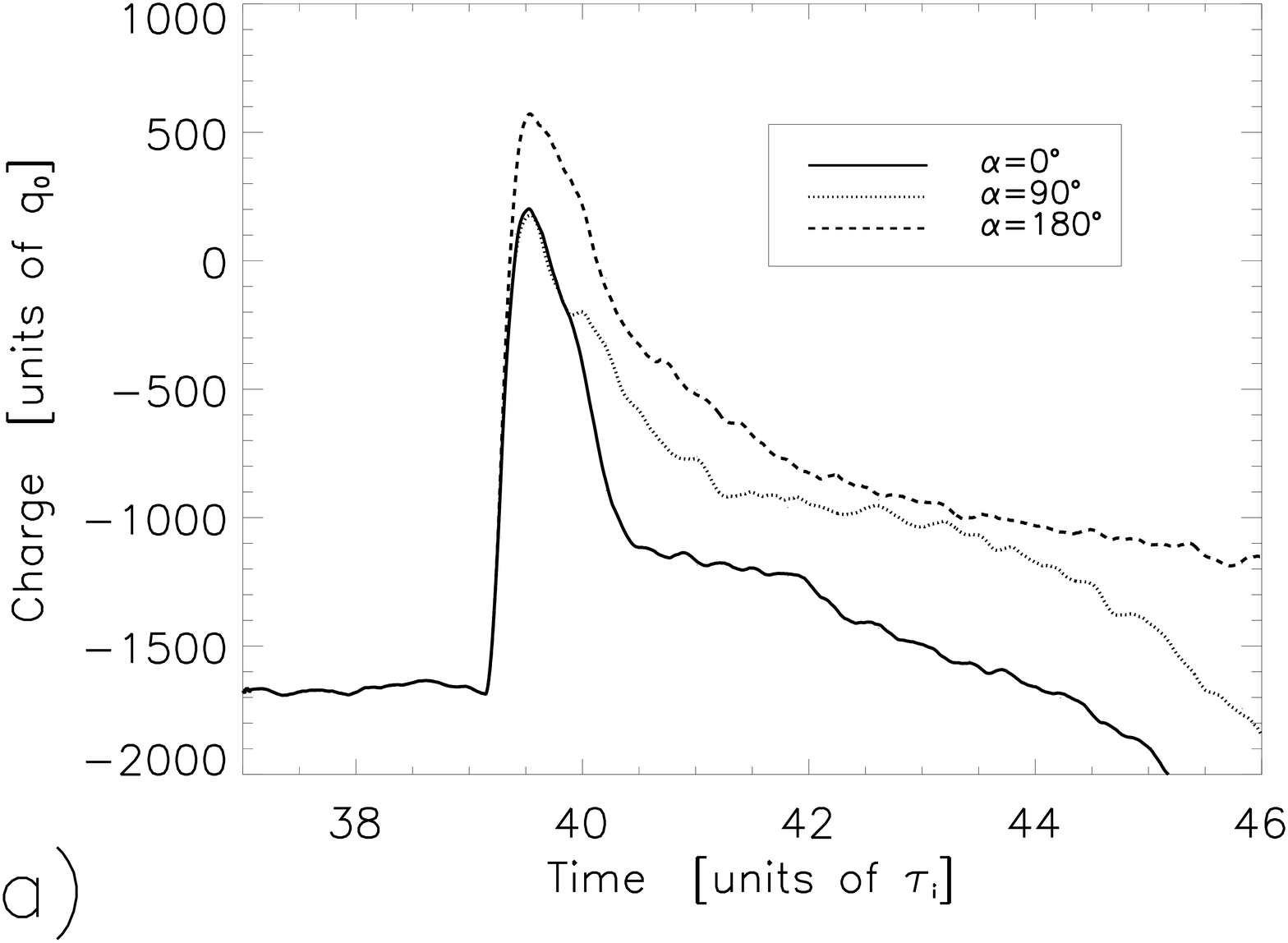}
\includegraphics[width=0.59\columnwidth]{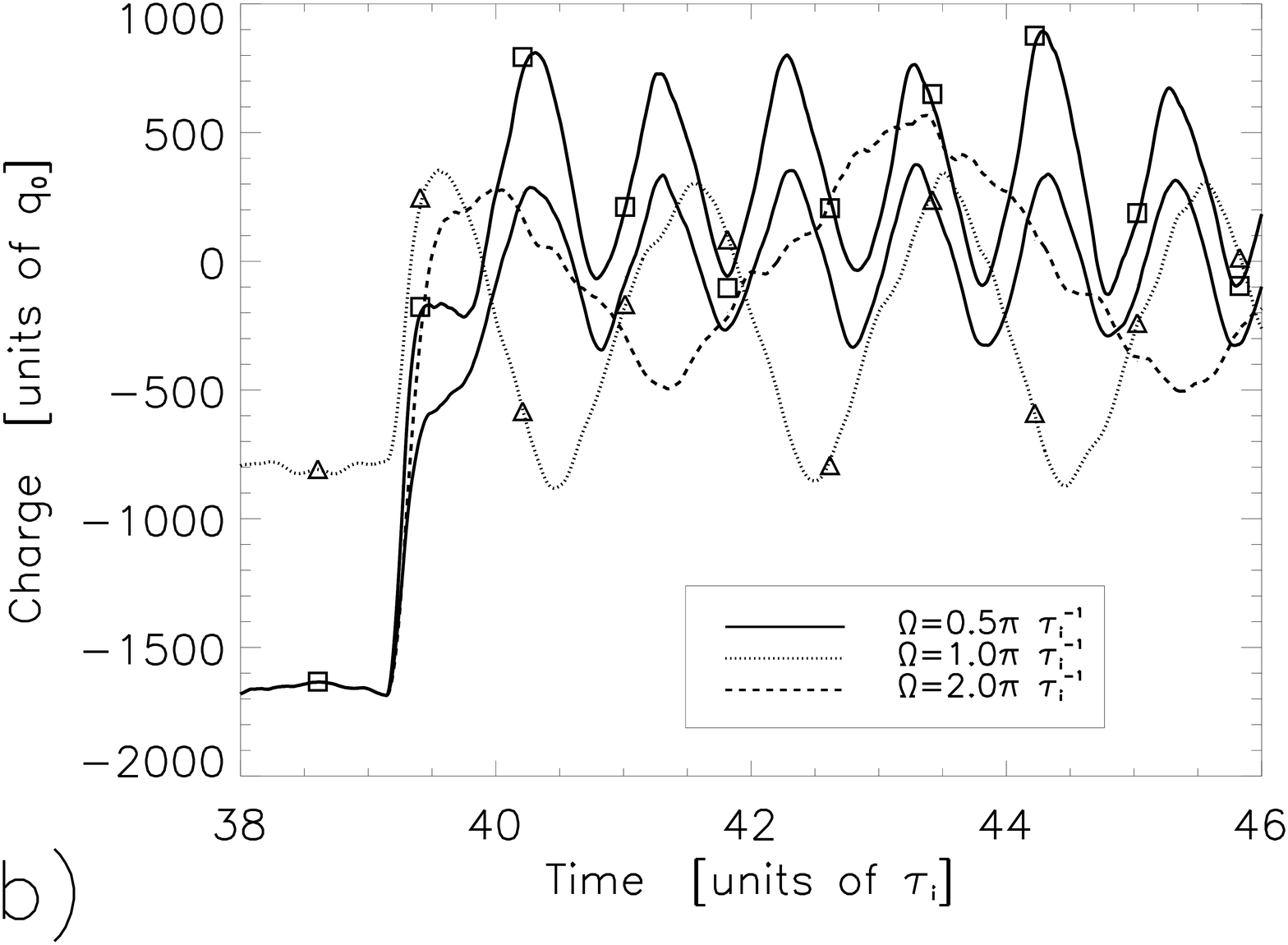}
\caption{The total charge on an insulating dust grain rotating with an angle of $\pi$ over one ion plasma period $\tau_i$ (a) and continuously rotating after the start of photo\-emission (b). Continuous radiation with $\Phi_{h\nu} = 1.25 \times 10^{19}~\mathrm{m^{-2}s^{-1}}$ (squares correspond to $\Phi_{h\nu} = 2.5 \times 10^{19}~\mathrm{m^{-2}s^{-1}}$) is switched on at $t=39\tau_i$. Triangles correspond to the dust grain spinning throughout the whole simulation. $\alpha=0^{\circ}$ for $\Omega=\pi$, $\alpha=90^{\circ}$ for $\Omega=0.5\pi$, and $\alpha=180^{\circ}$ for $\Omega=2\pi$ in units of $\tau_i^{-1}$. The results are smoothed with a moving box average filter for presentation.}
\label{fig:rotation}
\end{center}
\end{figure}

Continuous rotation by an angle of $\pi$ within one ion plasma period significantly modifies the charging of the grain, see Fig.~\ref{fig:rotation} (a). The rotation of a grain redistributes the charge on the dust grain surface, and lowers the  total charge on the grain. After arresting the grain rotation, the charge becomes more negative for $\alpha=0^{\circ}$ and $\alpha=90^{\circ}$, while for $\alpha=180^{\circ}$ it can saturate with a quadrupole moment in the surface charge distribution. 

With steady state rotation, the total charge oscillates in time with the mean charge value lower than on the conducting grain with the corresponding parameters for the photon flux. The period of oscillations depends on the angular velocity of the grain, see Fig.~\ref{fig:rotation}b). The oscillations are not observed for very slow angular velocities. There is little difference in the grain charging characteristics for different starts of the rotation of the grain. For a grain spinning throughout the whole simulation, the total charge before the onset of  photo\-emission is less negative than on a stationary grain. This is due to the charge redistribution, which prevents the development of a strong electric dipole moment. However, with photo\-ionization, the charging characteristics are similar to the case when the dust grain starts spinning after the radiation onset.
The charge redistribution on spinning grains tilts the electric dipole moment on insulating grains. The strength of the electric dipole moment oscillates together with the total charge on the grain. Simultaneously, the wake becomes asymmetric and its size oscillates in time.

\subsection{Pulsed radiation}

\begin{figure}[!htb]
\begin{center}
\includegraphics[width=0.59\columnwidth]{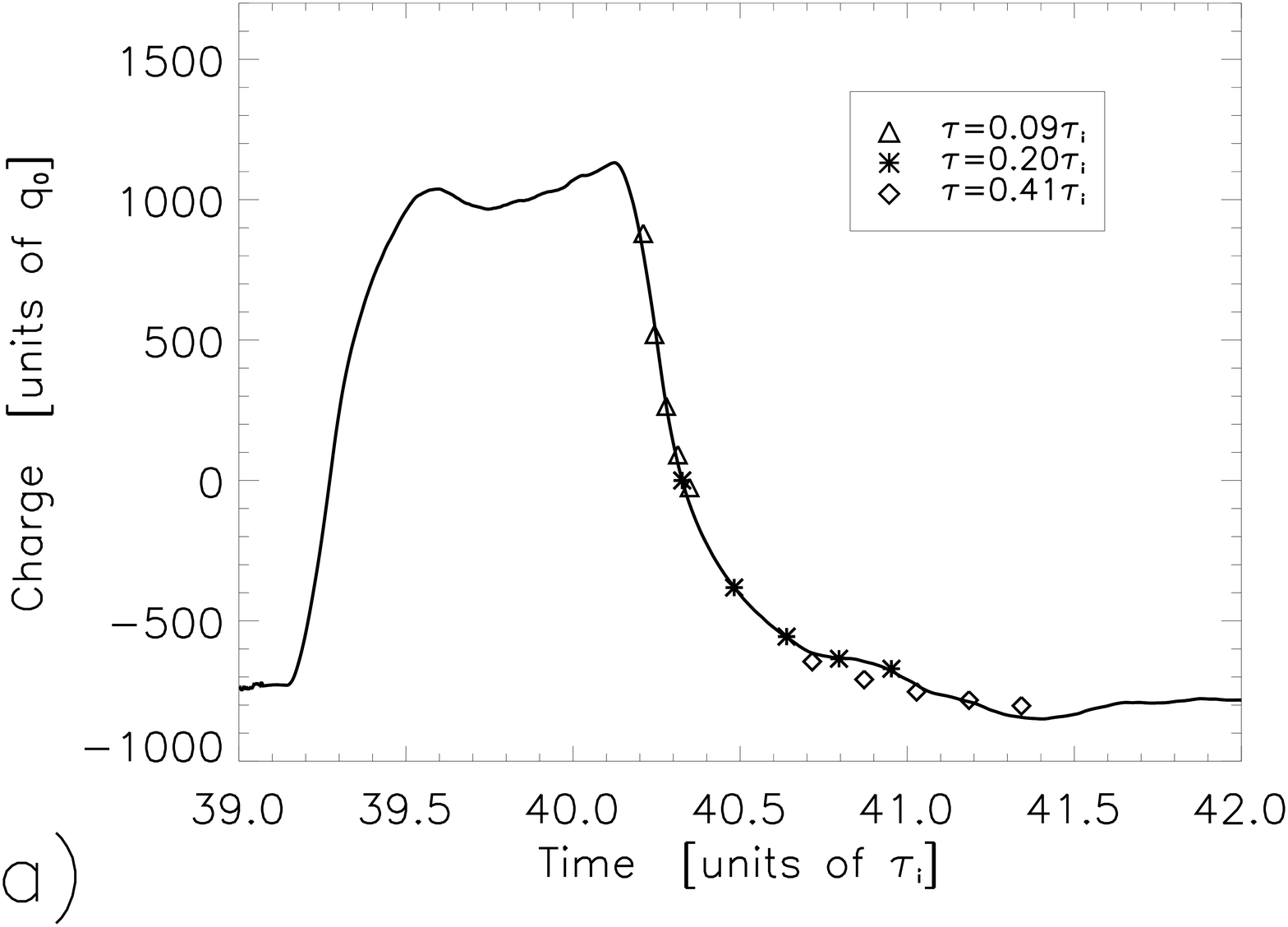}
\includegraphics[width=0.59\columnwidth]{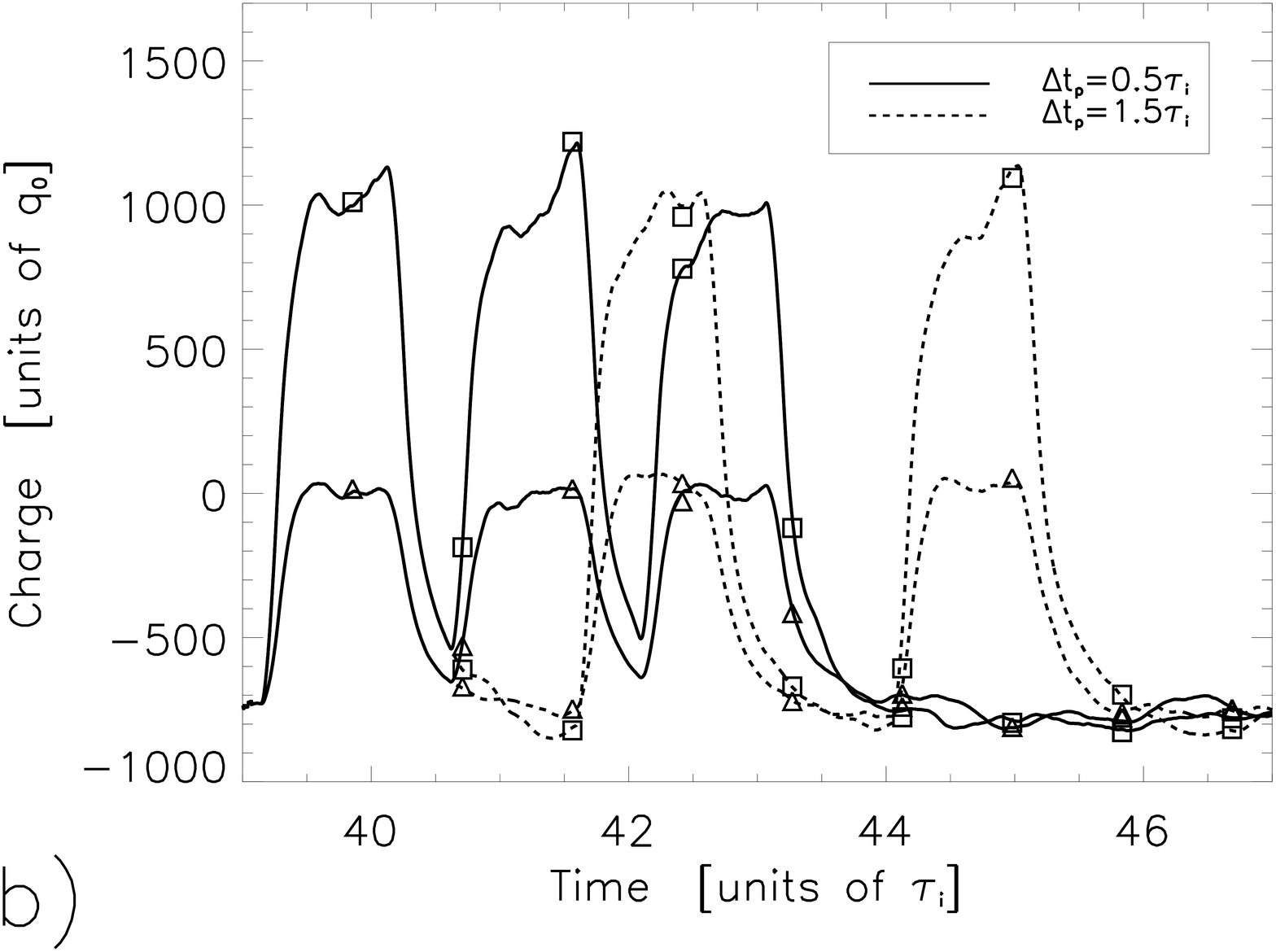}
\caption{The total charge on a conducting dust grain exposed to a single radiation pulse (a) and to a pulse series with different  time intervals between pulses $\Delta t_p$ (b). In plot (a) different symbols represent regularly spaced data points corresponding to local exponential fits with different dime constants $\tau$. In plot (b)  triangles correspond to $\Phi_{h\nu} = 0.5 \times 10^{19}~\mathrm{m^{-2}s^{-1}}$ and squares to $\Phi_{h\nu} = 2.5 \times 10^{19}~\mathrm{m^{-2}s^{-1}}$. In both cases $\alpha=0^{\circ}$, and $E_{h\nu}=5.5~\mathrm{eV}$. The results are smoothed with a moving box average filter for presentation.}
\label{fig:pulse_cond}
\end{center}
\end{figure}

The charge on the conducting grain exposed to a radiation pulse is more positive during the illumination. After the pulse, the charge recovers to the previous value (before the pulse) within approximately one ion plasma period. The charge recovery is initially fast and then continues at a slower rate. Initially, the charge recovery can be well approximated by an exponential function of the form $q=q_0\exp[-t/\tau]$, with the time constant $\tau=3.45 \times 10^{-9} \mathrm{s}$ for $\Phi_{h\nu} = 2.5 \times 10^{19}~\mathrm{m^{-2}s^{-1}}$, and $\tau=4.56 \times 10^{-9} \mathrm{s}$ for $\Phi_{h\nu} = 0.5 \times 10^{19}~\mathrm{m^{-2}s^{-1}}$. These time constants are comparable with the electron plasma period  $\tau_e=3.53 \times 10^{-9} \mathrm{s}$, which suggests that initially the charge recovery is primarily due to electrons. The time constant  for  $\Phi_{h\nu} = 0.5 \times 10^{19}~\mathrm{m^{-2}s^{-1}}$ is larger than $\tau_e$ because the maximum charge is close to zero in this case, and the ions contribute initially to the charge recovery.  After a time interval of $2 \tau_e$, the time constant $\tau$ is larger and reaches $\tau \approx 0.5 \tau_i$ at the end of the recovery for both cases. 
The charging is shown in Fig.~\ref{fig:pulse_cond}a) together with points corresponding to the exponential fits. For clarity of presentation, we do not show continuous exponential fits, but only regularly spaced points corresponding to the locally fitted curves.  Approximately one ion plasma period after the switch off, a small overshoot in the charging characteristic is observed for higher photon energies. The charging at given photon fluxes depends only little on the photon incident angles $\alpha$.

The charging after a series of three pulses is similar to what is found for a single pulse with the relevant photon flux and energy. Each pulse corresponds to a peak in the charging characteristics in Fig.~\ref{fig:pulse_cond}b). The height of each peak does not change much with the time interval between the pulses. The trough is less negative for time intervals between the pulses that are shorter than the charge recovery time.

\begin{figure}[!htb]
\begin{center}
\includegraphics[width=0.59\columnwidth]{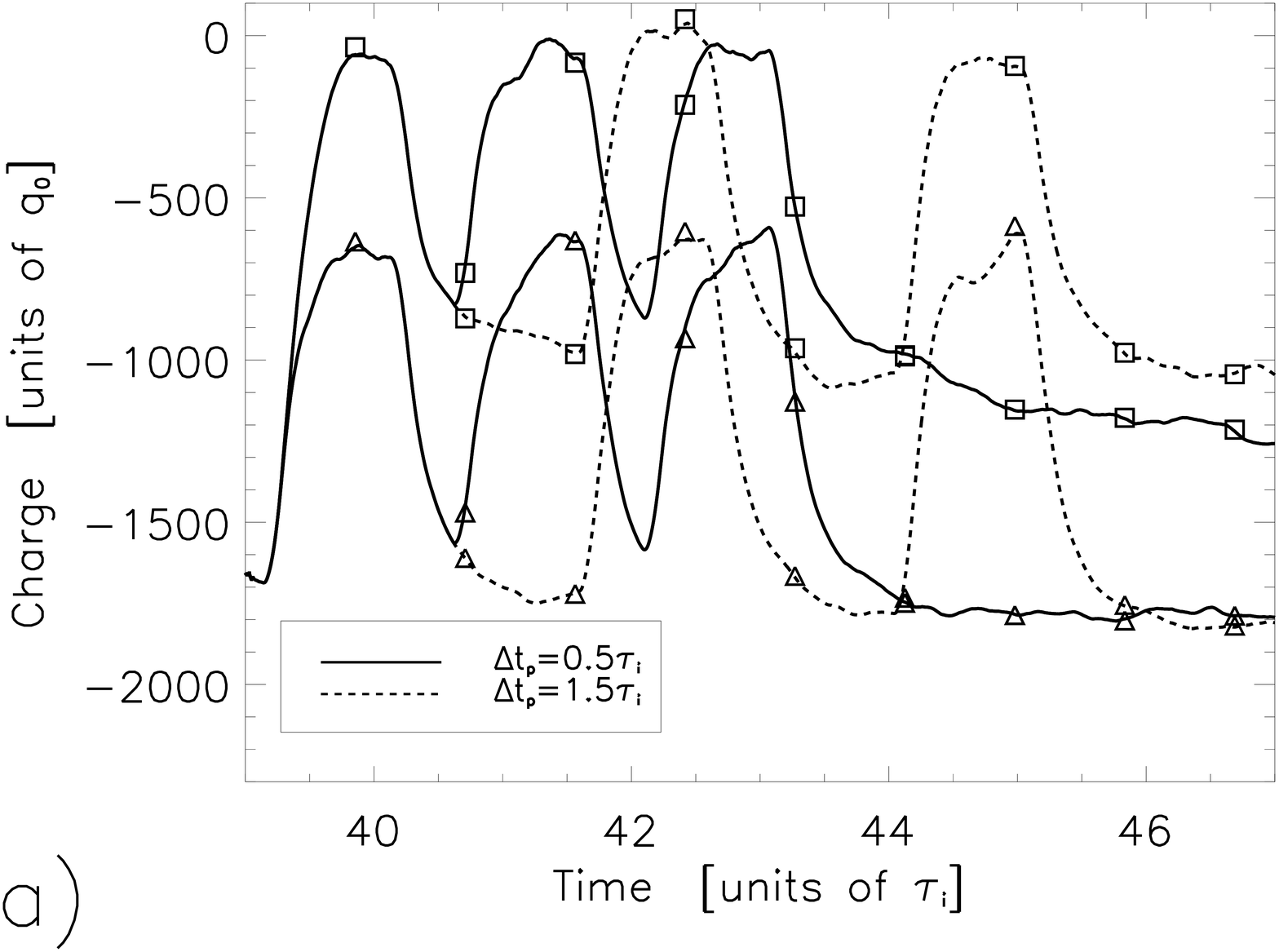}
\includegraphics[width=0.59\columnwidth]{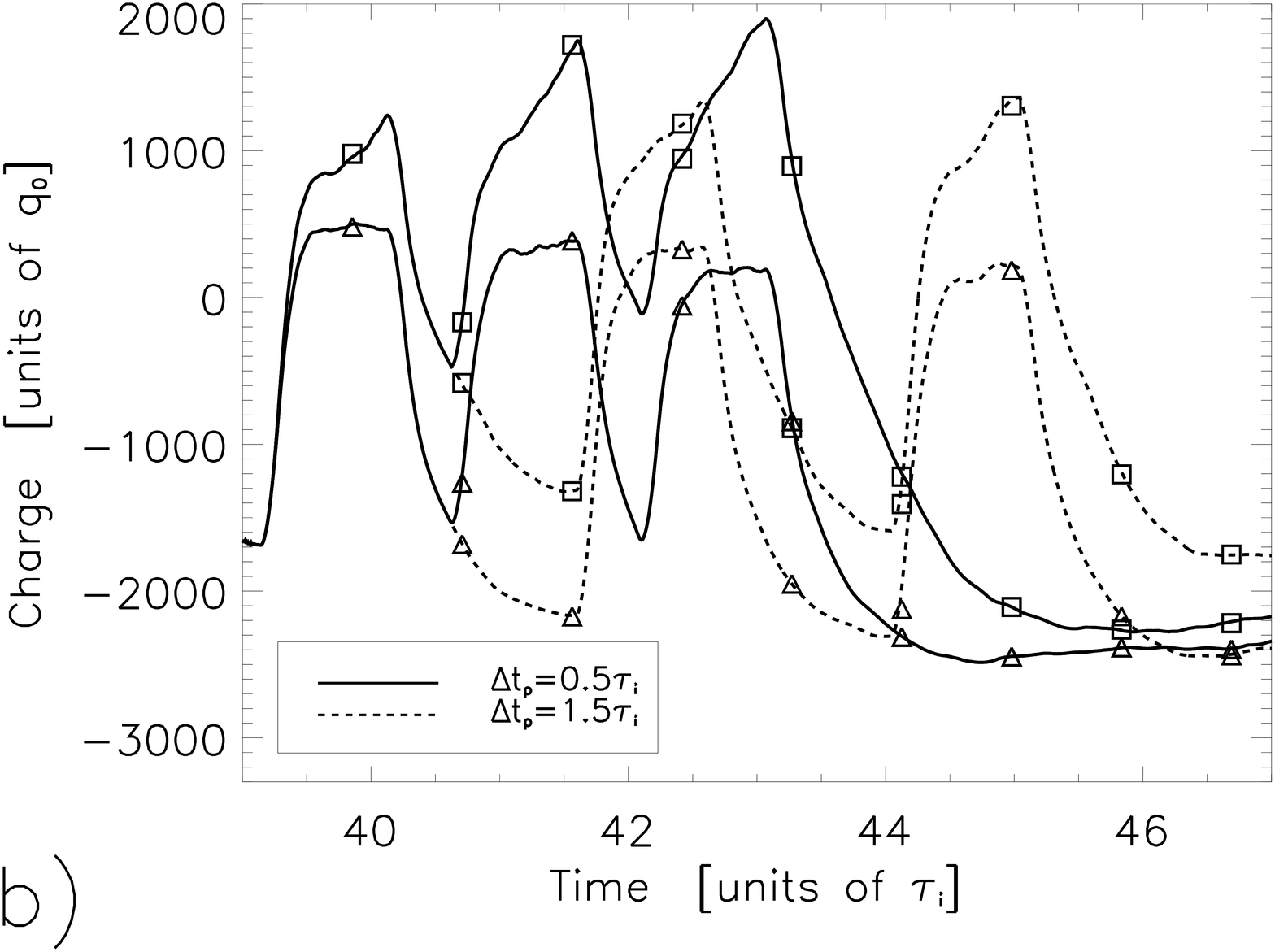}
\caption{The total charge on an insulating dust grain exposed to the pulsed radiation as a function of time for different time intervals between pulses $\Delta t_p$ and different photon fluxes: $\Phi_{h\nu} = 0.5 \times 10^{19}~\mathrm{m^{-2}s^{-1}}$ (a) and $\Phi_{h\nu} = 2.5 \times 10^{19}~\mathrm{m^{-2}s^{-1}}$ (b). Squares correspond to $\alpha=180^{\circ}$, triangles to $\alpha=0^{\circ}$. The photon energy is $E_{h\nu}=11.0~\mathrm{eV}$. The results are smoothed with a moving box average filter for presentation.}
\label{fig:pulse_ins}
\end{center}
\end{figure}

\begin{figure}[!htb]
\begin{center}
\includegraphics[width=0.9\columnwidth]{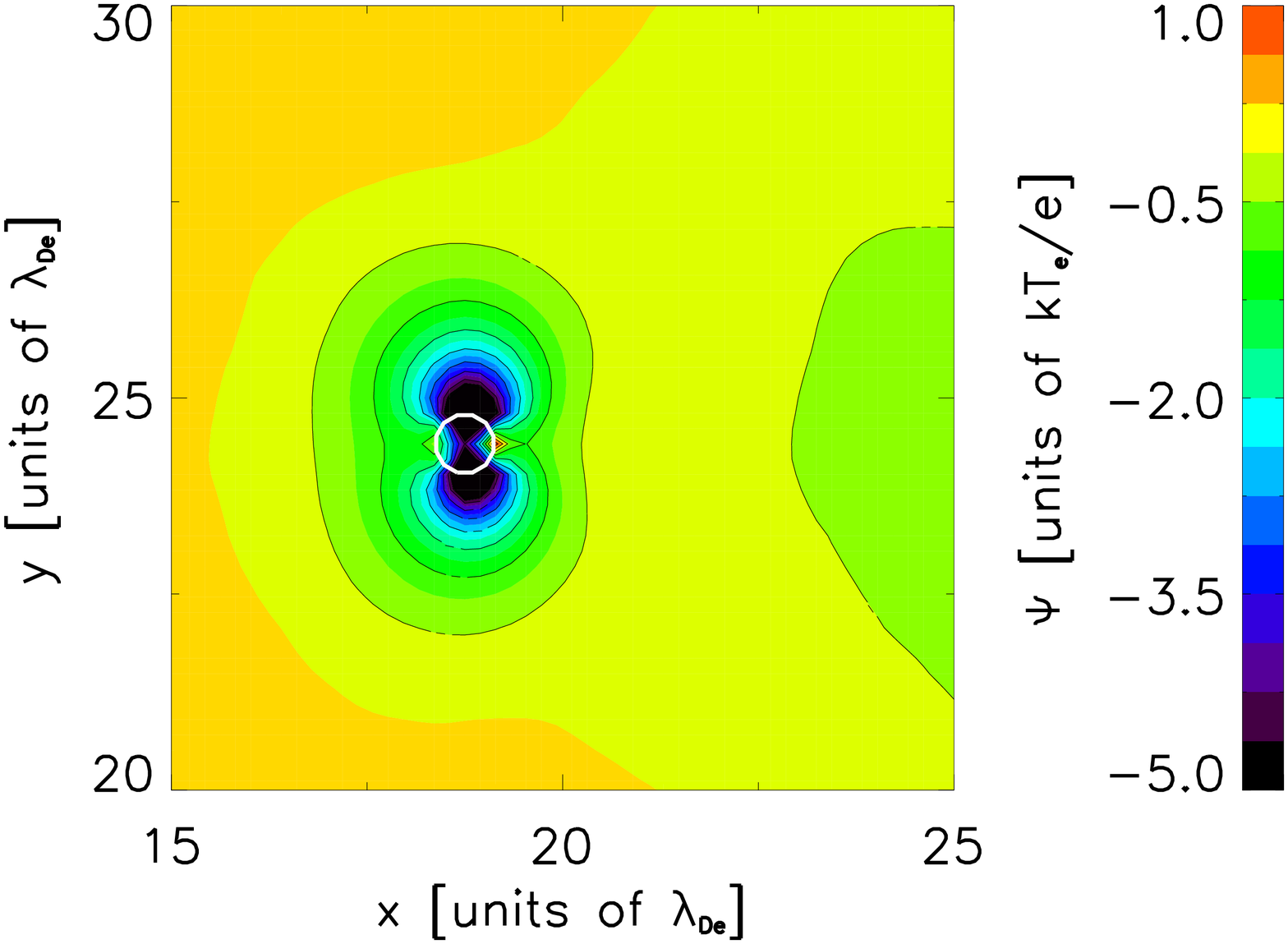}
\caption{The potential around an insulating dust grain after a series of pulses with $\Delta t_p=1.0\tau_i$. $\Phi_{h\nu} = 2.5 \times 10^{19}~\mathrm{m^{-2}s^{-1}}$, $E_{h\nu}=11.0~\mathrm{eV}$, and $\alpha=180^{\circ}$. The minimum in the potential is $\Psi=-12.2$ in units of $kT_e/e$ on the surface of the dust grain. Potentials lower than $\Psi=-5~kT_e/e$ are coloured black.}
\label{fig:quadrupole}
\end{center}
\end{figure}

\begin{figure}[!htb]
\begin{center}
\includegraphics[width=0.9\columnwidth]{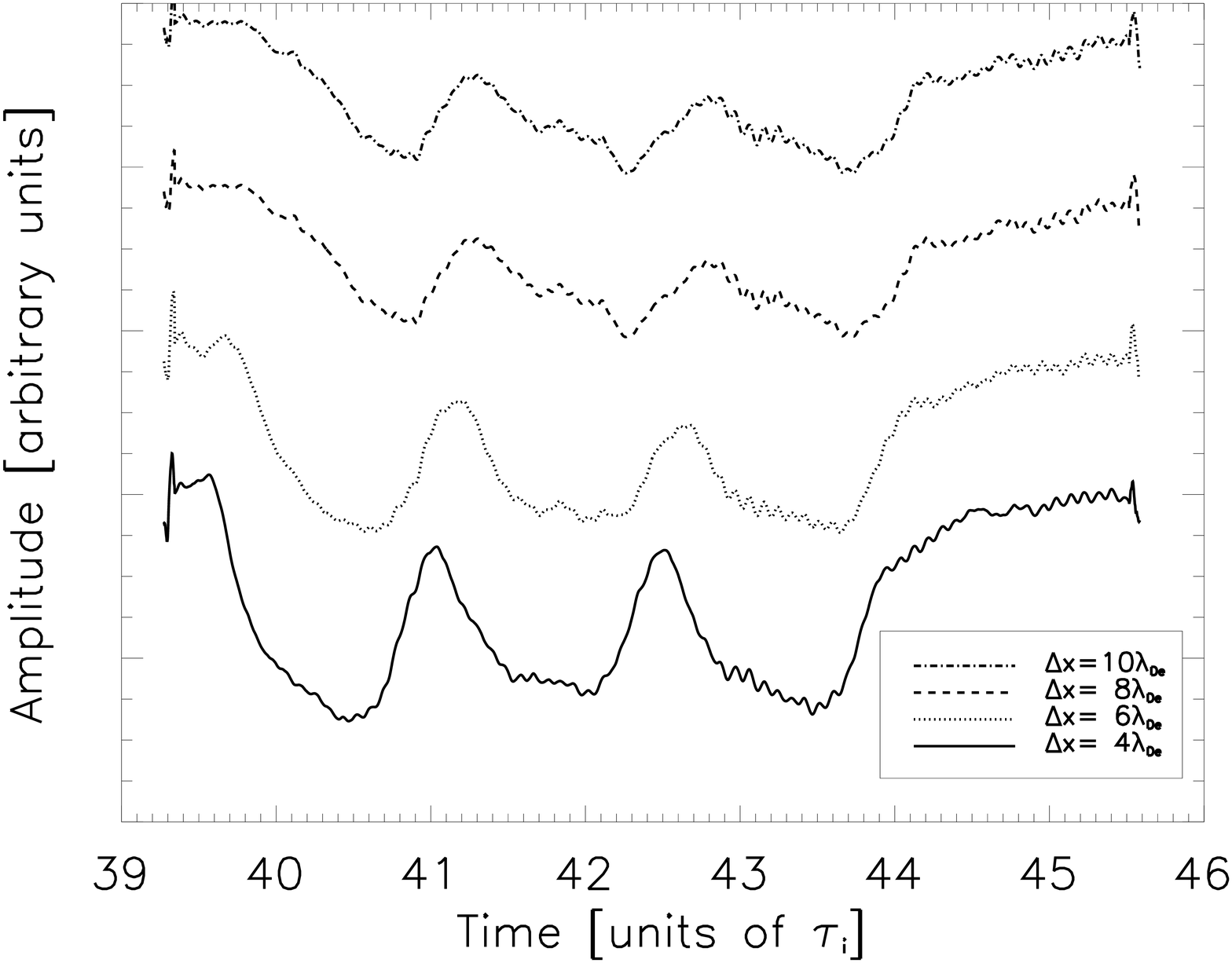}
\caption{The potential variations at different distances $\Delta x$ from the rear of the conducting grain for $y=25.8$ in units of $\lambda_{De}$ exposed to the radiation pulses as a function of time. $\Phi_{h\nu} = 2.5 \times 10^{19}~\mathrm{m^{-2}s^{-1}}$, $E_{h\nu}=5.0~\mathrm{eV}$, and $\alpha=180^{\circ}$.}
\label{fig:oscillations}
\end{center}
\end{figure}

The electrostatic potential around the conducting dust grain exposed to  radiation pulses is polarized as in the case of continuous radiation. During the pulses, the potential behind the dust grain is negative, and resembles the case of the conducting grain with continuous radiation, see Fig.~\ref{fig:polarization_c}. This region remains negatively charged also between the pulses. Within a time interval of approximately $1.5\tau_i$ after the last pulse, the positive potential region in the grain wake is rebuilt: first in the vicinity, and then further away from the grain. At the same time, the region with net negative charge becomes less pronounced and moves further downstream from the dust grain, slower than the ion drift speed. During the pulses, the wake potential in the vicinity of the grain oscillates with the frequency of the pulses, see Fig.~\ref{fig:oscillations}. These oscillations propagate into the wake, but are heavily damped further away from the dust grain, and diminish after the last pulse.

The ion density behind a conducting dust grain is being enhanced between the pulses. This process is terminated by the start of a successive pulse, and the enhanced ion density regions move away from the dust grain in the ion drift direction, but at a lower speed. 
The full recovery of the ion focus occurs after approximately one  ion plasma period from the last pulse. The ion density wake in front of the grain is located closer to the grain for $\alpha=180^{\circ}$ than for $\alpha=0^{\circ}$. There is also a depletion in the electron density in the region corresponding to the ion wake originating from a positively charged grain. After the switch-off, photo\-electrons are rapidly redistributed, but the depletion in the region corresponding to the wake remains until the wake is filled with ions.


An insulating dust grain exposed to a single radiation pulse has charging characteristics similar to the conductor only for low photon fluxes and $\alpha=0^{\circ}$, when the total grain charge remains negative, see Fig.~\ref{fig:pulse_ins}a). For the low photon flux and $\alpha=180^{\circ}$, the total charge on a dust grain is low during the pulses. In this case, the charge does not recover within one ion plasma period after the pulse, but it reaches half of the charge value before the onset of radiation. Longer simulations show that the recovery time is approximately 20 ion plasma periods in this case. 

For high photon fluxes, the charge during the pulse is positive, see Fig.~\ref{fig:pulse_ins}b).  After a single pulse with  $\alpha=0^{\circ}$ the charge is more negative than before the pulse, and less negative when $\alpha=180^{\circ}$. The total charge in the trough between subsequent pulses is getting more negative for $\alpha=0^{\circ}$, while for $\alpha=180^{\circ}$ it is more negative for longer time intervals between pulses, and less negative for shorter intervals. In all insulating cases, the total charge after a series of pulses can be more negative than after a corresponding single pulse, and the full charge recovery takes usually several plasma periods. 


The radiation pulses modify the potential and density patterns in the vicinity of the grain. For photons with  $\alpha=0^{\circ}$, we observe an  enhancement in the electric dipole moment on the dust grain.  For photons with $\alpha=180^{\circ}$, the positive surface charge is accumulated on the front and rear sides of the dust grain. During such pulses, the electric dipole moment is parallel to the ion flow (antiparallel to the photons direction), and after the pulses, the quadrupole moment in the surface charge distribution develops, see Fig.~\ref{fig:quadrupole}. The quadrupole moment diminishes in time, faster for low photon fluxes, and the dipole moment in the surface charge distribution as well as the electrostatic potential distribution are recovered.

In the wake, the ion focusing is rebuilt behind the dust grain after the radiation pulses.  For $\alpha=0^{\circ}$, this region is similar to the one before the pulses, while for $\alpha=180^{\circ}$, at the same time instances it is wider and weaker for low fluxes, and spatially narrower with stronger focusing for high fluxes. Electrons are Boltzmann distributed after the pulses. 

\section{Discussion}
Photo\-emission provides an electron source on the irradiated side of the grain and modifies the dust grain charge. For sufficiently high photon fluxes, the charge on the conducting grain becomes positive and saturates within one ion plasma period. Positively charged conducting grains slow down and deflect flowing ions. As a result, a region of enhanced ion density forms in front of the grain, while behind the grain a substantial wake in the ion density is formed, see again Fig.~\ref{fig:uv_ionwake_c}. The wake in the ion density scales with the photon energy and flux, being larger for higher fluxes and energies. Hence, the wake size is proportional to the charge on the grain. Photo\-electrons with energies higher or comparable to the electron thermal velocity can easily be lost on the grain surface, while with higher energies they are more likely to escape the trapping potential of the grain. This together with the photo\-emission rate, which is proportional to the flux, explains the development of more positive charge on the grain for high energetic photons and high fluxes \cite{Miloch_Vladimirov_2008}. The angle between incoming photons and plasma flow direction has little effect on the potential distribution around conducting grains. Photo\-electrons contribute in neutralizing enhanced ion density regions. For this reason, the region of enhanced ion density is located closer to the front of the grain for $\alpha=0^{\circ}$, when the grain charge can be effectively shielded by the photo\-electrons, and further away for $\alpha=180^{\circ}$, when the photo\-electrons are produced on the shadow side. The electrons penetrate into the ion wake, due to their high mobility. The resulting imbalance between ion and electron densities in the vicinity of the grain leads to polarization of the plasma, see again Fig.~\ref{fig:polarization_c}. This allows for strong interactions between many positively charged grains in flowing plasmas. The electrons are no longer Boltzmann distributed.

To calculate the charge on positively charged conducting grains, we use a two-dimensional capacitance model \cite{Miloch_Pecseli_Trulsen_2007}. In (\ref{q_theory}) we use the simulation results for the floating potential $\Psi_{fl}$. For photon fluxes of $\Phi_{h \nu}=1.25 \times 10^{19}~\mathrm{m^{-2}s^{-1}}$, the calculated charge is $q_t \approx 250 q_0$ for both photon energies. This result is close to the data shown in Table \ref{tab:uv_charging_c}. For $\Phi_{h \nu}=2.5 \times 10^{19}~\mathrm{m^{-2}s^{-1}}$ we have  $q_t \approx 410 q_0$ for $E_{h \nu}= 4.8~\mathrm{eV}$, and  $q_t \approx 732 q_0$ for $E_{h \nu}= 5.5~\mathrm{eV}$, which is lower than the simulation results. However, if  in equation (\ref{q_theory}) we formally substitute $\lambda_{D}$ by $\lambda_{De}$ for  $\Phi_{h \nu}=2.5 \times 10^{19}~\mathrm{m^{-2}s^{-1}}$, then the results are close to the simulation results: $q_t \approx 734 q_0$ for $E_{h \nu}= 4.8~\mathrm{eV}$, and  $q_t \approx 1312 q_0$ for $E_{h \nu}= 5.5~\mathrm{eV}$. This suggests that for lower photon fluxes, and low positive potentials of the grain, the ions can effectively contribute to the shielding of the grain, while for more positive potentials, the grain potential is predominantly shielded by electrons.

The analytical solution for the floating potential is in a good agreement with the simulation results for low energy photons. For high energy photons the analytical calculations give more positive potentials than obtained from the simulations. This is due to  thermalization of the photo\-electrons. The temperature of low energy photo\-electrons is higher than the mean temperature of the background electrons, but the corresponding velocities are still within the thermal spread of the background electrons. High energy photo\-electrons are effectively slowed down by the grain and interact with background electrons. We find that the average rate of deceleration of the high energy photo\-electrons is $25\%$ of their initial energy. With this correction for high energy photons, the analytical calculations for the floating potential give values close to the numerical results.

An electric dipole moment develops on insulating grains due to the photo\-emission. It is oriented antiparallel to the photon direction. Neither the electric dipole moment nor the charge saturate on such grains within the simulation time. For $\alpha=180^{\circ}$ the charge is more positive in time, while for other angles of incidence it recovers to negative values. For $\alpha=180^{\circ}$ both rear and front of the grain are positively charged. The depletion in the ion density behind the grain does not allow electrons to neutralize the rear charge. With increasing charge on the rear of the grain, a weak quadrupole moment develops (with negatively charged grain sides tangential to the flow), and the total charge increases towards positive values. For other angles, the non-irradiated side of the grain becomes more negative when the photo\-electric current exceeds the electron current.
Therefore, a depletion of the wake and a recovery of the ion focusing region are observed, see again Figs.~\ref{fig:polarization_i} and \ref{fig:uv_ionwake_i}. The surface charge distribution on insulating grains for $\alpha=90^{\circ}$ leads to more pronounced asymmetries in the ion density than on conducting  grains. The electrons are no longer Boltzmann distributed also in the case of insulators, but a dominant contribution here is due to the electric dipole development. Photo\-electrons are absorbed by the high positive charge on the dust grain, and they neutralize enhanced ion density regions in front of the grain. 

Rotation of the insulating dust grain redistributes the charge on the grain surface. Without photo\-emission but retaining the directed ion flow, the total charge on fast spinning grains becomes less negative. The electric dipole moment on the surface diminishes and the value of the total charge is similar to the conducting case. With photo\-emission, the electric dipole moment on the spinning dust grain is still present for the photon fluxes considered in this work. It is tilted by an angle with respect to the direction of radiation, and the positive charge on the irradiated side of the dust grain is being neutralized when it reaches the shadow side. Due to the depletion of the wake in the ion density, the total charge oscillates on fast spinning grains, see Fig.~\ref{fig:oscillations}. 
With the spinning grain, the symmetry in the ion wake is destroyed near the grain surface. For sufficiently fast rotation, the redistribution of the negative charge bends ion trajectories and leads to wake erosion. This process continues until the ion density is rebuilt in the vicinity of the grain and the region of reduced density is detached from the grain. The electron density in the wake increases, and so does the total electron current to the grain. When the charge becomes less positive, and the electron current to the grain decreases, photo\-emission leads to the formation of the new wake in the ion density. 
The closing of the wake and the resulting oscillating total charge will occur only if the erosion of the ion wake is substantial. If the rotation of the grain is slow and the photo\-emission rate is high, the wake will not close and detach. 

The present discussion considers grains with spherical, cylindrical and oblate shapes.  
Irregularities on the dust grain surface can lead to variations in the surface charge distribution and the wake can be perturbed also for smaller angular velocities of the grain. It is also noted, that spherical grains with inhomogeneous surface properties will spin due to angular momentum transfer from ions \cite{Tsytovich_Vladimirov_2004}. Such inhomogeneities can also lead to complicated surface charge distributions, as illustrated elsewhere \cite{Miloch_Pecseli_Trulsen_2007}.

Due to the high inertia of the dust grains, the rotation of the grains will be slow in most experiments 
 \cite{Piel_Melzer_2002}. Hence, irregularities in the dust grain surface that give rise to a complicated surface charge distribution will be the main factor for the charge saturation on the insulating grain in the presence of directed radiation. This will be valid also for slowly spinning grains.
On the other hand, the surface charge distribution on the insulating dust grain exposed to directed radiation will lead to strong electric fields within the grain. This effect will be more pronounced on grains with surface irregularities. The irregularities will eventually be destroyed by strong electric fields. This will be valid also for stationary plasmas and is similar to the sterilization and destruction of bacteria by means of plasma used as a source of UV radiation \cite{McDonald_Curry_2002, Laroussi_Mendis_2003}. 

Photo\-emission provides a method for controlling the charge on conducting grains both in vacuum \cite{Sickafoose_Colwell_2000} and in plasma \cite{Miloch_Vladimirov_2008}. For perfectly insulating grains, the total charge saturates only when the total grain charge remains negative, but it depends on the photon incidence angle. The charge should also saturate for slow spinning insulators due to surface irregularities. It oscillates for fast spinning insulating grains, however.

After pulses of radiation, the charge recovers to the value from before the onset of radiation. The recovery is initially mainly due to electrons, and then both electrons and ions. The charging can be approximated by an exponential function with the time constant $\tau$ that is initially comparable with the electron plasma period and then larger. This is in agreement with the previous results from dust grain charging simulations \cite{Miloch_Pecseli_Trulsen_2007}. The charge on conducting grains recovers within one ion plasma period. The small overshoot in the charging characteristics at approximately one ion plasma period after the switch off can be attributed to the ion response due to the reduced ion mass \cite{Miloch_Pecseli_Trulsen_2007}, and the formation of the ion focus. For insulators, the charge recovery may take up to 20 ion plasma periods. This is due to a complicated surface charge distribution on the grain. 
For $\alpha=180^{\circ}$, the positive charge on the rear of the grain is reduced slowly after the pulse because of the reduced density in the wake, while the front side of the grain is positively charged by the ion flow. A quadrupole moment is present in the surface charge distribution for several ion plasma periods after a pulse. Consistently, the charge between and after the pulses is less negative for $\alpha=180^{\circ}$ than for other angles. For other angles, the surface charge distribution leads to a more negative total charge after the switch off as compared to the dust grain charge previous to photo\-emission. 
 
The photon fluxes considered for conductors in this work, can be achieved by commercially produced sources of UV radiation (e.g., low pressure mercury lamps) \cite{McDonald_Curry_2002}. In case of lamps it would be necessary to collimate the light, but in case of UV lasers the energies of $E_{h\nu} \in (4.8,7.2)\mathrm{eV}$, corresponding to $\lambda \in (172,258)\mathrm{nm}$, can be achieved for instance by excimer lasers used in photolithography \cite{Ewing_2000}. The photon energies in the range $E_{h\nu} \in (10.3,12.7)\mathrm{eV}$, corresponding to $\lambda \in (97,120) \mathrm{nm}$, are more difficult to obtain, but can still be achieved for instance by free electron lasers \cite{Neil_Meriminga_2002}. In our work we assumed the work function for the insulator to be $W=10~\mathrm{eV}$, which is similar to values from experiments with ice \cite{Westley_Baragiola_1995, Baragiola_Vidal_2003}. However, insulating grains can have lower work functions. The work function of pure ice is $W \approx 8.7~\mathrm{eV}$, and can be significantly lower if the ice contains impurities, as it is expected in the atmosphere \cite{Klumov_Morfill_2005}. Sodium silicate glass can have work function as low as $W=6~\mathrm{eV}$ \cite{Vishnakov_Trukhin_1991}. Therefore, on such insulators similar effects, as demonstrated in this study, should be achieved by lower photon energies. 

By illuminating a grain using short pulsed lasers, it is possible to modify the charge on the grain and excite potential oscillations in the wake. The other dust grain, if located in the wake, will experience oscillations. Since the charge on the illuminated grain is determined by photo\-emission, the motion of the particle would provide non-invasive diagnostics for measuring the charge on the grain located in the wake of the illuminated one. On the other hand, continuous illumination of the grain placed in the wake will fix the grain charge and allow for accurate study of the wake of other grain.

The results presented here does not depend on the ion to electron mass ratio, except for the saturation charge on the dust grain without photo\-emission and the charging rate. However, the charge on the grain with photo\-emission does not change with the ion to electron mass ratio. This is because the photo\-emission current does not change with the ion to electron mass ratio, and the ion current to positively charged grains is negligible. 

There are certain limitations of our model. The radiation is assumed to be unidirectional and photo\-electrons to be monoenergetic. In many problems in laboratory plasmas, UV radiation will be more isotropic due to the plasma glow and light scattering, while photo\-electron energies will be statistically distributed. Isotropic radiation will cause more homogeneous distributions of the photo\-electons and the grain surface charge. We considered perfectly insulating and perfectly conducting dust grains. Finite conductivity due to impurities and resistivity can modify the results, especially for insulators. These issues were not considered in this work.

\section{Conclusions}
The results from numerical simulations of charging of isolated dust grains in flowing plasmas with photo\-emission were presented for perfectly insulating and perfectly conducting grains. By means of photo\-emission, the total charge on a conducting grain can be effectively controlled. The charge control on insulating grains is more difficult since no charge saturation is observed for stationary grains, and the charging characteristics depend on the angle between the incident photons and the plasma flow. For insulating grains, surface charge irregularities and rotation of the dust grain can redistribute the surface charge on the grain. Fast spinning of the grain results in oscillations of the value of the total grain charge and the density wake behind it.

During photo\-emission, the electrons are non-Boltzmann distributed in the vicinity of the grain. This makes a theoretical analysis of the problem difficult. The plasma is polarized in the vicinity of the grain, which can give rise to strong interactions between many dust grains. For insulators the interactions are controlled by a strong electric dipole moment on the surface, antiparallel to the direction of radiation. After pulses of radiation, the charge, density and potential distributions recover to the conditions from before the photo\-emission. The recovery takes approximately one ion plasma period for conducting grains, and several times longer for insulating grains, due to the complicated charge distributions on the dust grain surface. 

By a fine adjustment of the charge with the use of photo\-emission, the coagulation of the dust grains can be induced due to large relative fluctuations of the charge when the total charge on a grain is small. Both continuous and pulsed radiation should allow for non-invasive diagnostics of the charge and wake structure in dusty plasma experiments.

\ack
This work was in part supported by the Norwegian Research Council, NFR, and by the Australian Research Council, ARC. Two of the authors (WJM and HLP) wish to thank Dr. J{\o}rgen Schou  for useful discussions on the photo\-emission from insulating materials.


\section*{References}
\begin{thebibliography}{10}

\bibitem{Shukla_Mamun_2002}
Shukla P K and  Mamun A A. 2002
\newblock {\em Introduction to Dusty Plasmas}
\newblock (Bristol: Institute of Physics Publishing)

\bibitem{Vladimirov_Ostrikov_2005}
Vladimirov S V, Ostrikov K and Samarian A A 2005
\newblock {\em Physics and applications of complex plasmas}
\newblock (London: Imperial College Press)

\bibitem{Ishihara_2007}
Ishihara O 2007
\newblock Complex plasma: dusts in plasma
\newblock {\em J. Phys.} D: {\em Appl. Phys.} \textbf{40} R121--R147

\bibitem{Vladimirov_Nambu_1995}
Vladimirov S V and Nambu M 1995
\newblock Attraction of charged particulates in plasmas with finite flows
\newblock {\em Phys. Rev.} E \textbf{52(3)} R2172

\bibitem{Melzer_Schweigert_1996}
Melzer A, Schweigert V A, Schweigert I V, Homann A,  Peters S and Piel A 1996
\newblock Structure and stability of the plasma crystal
\newblock {\em Phys. Rev.} E \textbf{54} R46.

\bibitem{Ivlev_Morfill_1999}
Ivlev A V,  Morfill G, and Fortov, V E 1999
\newblock Potential of a dielectric particle in a flow of a collisionless plasma
\newblock {\em Phys. of Plasmas} \textbf{6}, 1415--20

\bibitem{Miloch_Pecseli_Trulsen_2008}
Miloch W J, Trulsen J and P{\'e}cseli H L 2008
\newblock Numerical studies of ion focusing behind macroscopic obstacles in a supersonic plasma flow
\newblock {\em Phys. Rev.} E \textbf{77} 056408

\bibitem{Melzer_schweigert_1999}
Melzer A, Schweigert V A and Piel A 1999
\newblock Transition from attractive to repulsive forces between dust molecules in a plasma sheath
\newblock {\em Phys. Rev. Lett.} \textbf{83} 3194--97

\bibitem{Maiorov_Vladimirov_2000}
Maiorov S A, Vladimirov S V and Cramer N F 2000
\newblock {Plasma kinetics around a dust grain in an ion flow}
\newblock {\em Phys. Rev.} E \textbf{63} 017401

\bibitem{Vladimirov_Maiorov_2003a}
Vladimirov S V, Maiorov S A and Cramer N F 2003
\newblock Kinetics of plasma flowing around two stationary dust grains
\newblock {\em Phys. Rev.} E \textbf{67} 016407

\bibitem{Hebner_Riley_2004}
Hebner G A and Riley M E 2004
\newblock Structure of the ion wakefield in dusty plasmas
\newblock {\em Phys. Rev.} E \textbf{69} 026405

\bibitem{Samarian_Vladimirov_2005}
Samarian A A, Vladimirov S V and James B W 2005
\newblock Wake-induced symmetry-breaking of dust particle arrangements in a  complex plasma
\newblock {\em JETP Lett.} \textbf{82} 758--762

\bibitem{Svenes_Troim_1994}
Svenes K R and Tr{\o}im J 1994
\newblock Laboratory simulation of vehicle-plasma interaction in low earth orbit
\newblock {\em Planet. Space Sci.}  \textbf{42} 81--94

\bibitem{Melandso_Goree_1995}
Melands{\o} F and Goree J 1995
\newblock Polarized supersonic plasma flow simulation for charged bodies such  as dust particles and spacecraft
\newblock {\em Phys. Rev.} E  \textbf{52} 5312

\bibitem{Horanyi_1996}
Hor\'{a}nyi M 1996
\newblock Charged dust dynamics in the solar system.
\newblock {\em Annu. Rev. Astron. Astrophys.} \textbf{34} 383--418

\bibitem{Hayakawa_Yamashita_1969}
Hayakawa S, Yamashita K and Yoshioka S 1969
\newblock Diffuse component of the cosmic far uv radiation and interstellar  dust grains.
\newblock {\em Astrophys. and Space Sci.} \textbf{5} 493--502

\bibitem{Wang_Horanyi_2007}
Wang X, Hora\'{a}nyi M, Sternovsky Z, Robertson S and Morfill G E 2007
\newblock A laboratory model of the lunar surface potential near boundaries between sunlit and shadowed regions.
\newblock {\em Geophys. Res. Lett.} \textbf{34} L16104

\bibitem{Sickafoose_Colwell_2000}
Sickafoose A A, Colwell J E, Hor{\'a}nyi M, and Robertson S 2000
\newblock Photoelectric charging of dust particles in vacuum
\newblock {\em Phys. Rev. Lett.}  \textbf{84} 6034--37

\bibitem{Weingartner_Draine_2001}
Weigartner J C and Draine B T 2001
\newblock Photoelectric emission from interstellar dust: grain charging and gas heating
\newblock {\em Astrophys. J. Suppl. Series} \textbf{134} 236--281

\bibitem{Klumov_Vladimirov_2005}
Klumov B A, Vladimirov S V and Morfill G E 2005
\newblock Features of dusty structures in the upper earth's atmosphere
\newblock {\em JETP Lett.} \textbf{82} 632--637

\bibitem{Klumov_Vladimirov_2007}
Klumov B A, Vladimirov S V and Morfill G E 2007
\newblock On the role of dust in the cometary plasma
\newblock {\em JETP Lett.} \textbf{85} 478

\bibitem{Schrittwieser_Ionita_2008}
R.~Schrittwieser R, Ionita C, Balan P, Gstrein R, Grulke O, Windisch T,
  Brandt C, Klinger T, Madani R, Amarandei G and  Sarma A K 2008
\newblock Laser-heated emissive plasma probe
\newblock {\em Rev. Sci. Instr.} \textbf{79} 083508

\bibitem{Rosenberg_Mendis_1995}
Rosenberg M and Mendis D A 1995
\newblock Uv-induced coulomb crystallization in a dusty gas
\newblock {\em IEEE Trans. Plasma Sci.} \textbf{23} 177

\bibitem{Rosenberg_Mendis_1996}
Rosenberg M, Mendis A and Sheehan D P 1996
\newblock Uv-induced coulomb crystallization of dust grains in high-presuure  gas
\newblock {\em IEEE Trans. Plasma Sci.} \textbf{24} 1422--29

\bibitem{Fortov_Nefedov_1998}
Fortov V E, Nefedov A P, Vaulina O S, Lipaev A M,  Molotkov V I,  Samarian A A,  Nikitskij V P, Ivanov A I,  Savin S F, Kalmykov A V,  Solov'ev A Y and Vinogradov P V 1998
\newblock Dusty plasma induced by solar radiation under microgravitational
  conditions: an experiment on board the mir orbiting space station
\newblock {\em J. Exp. and Theoret. Physics} \textbf{87} 1087--97

\bibitem{Samarian_Vaulina_2000}
Samarian A A and Vaulina O S 2000
\newblock Uv-induced coulomb structure in discharge plasma
\newblock {\em Phys. Lett.} A  \textbf{278} 146--151

\bibitem{Khrapak_Nefedov_1999}
Khrapak S A, Nefedov A P, Petrov O F and Vaulina O S 1999
\newblock Dynamical properties of random charge fluctuations in a dusty plasma  with different charging mechanisms
\newblock {\em Phys. Rev.} E  \textbf{59} 6017

\bibitem{Ostrikov_Yu_2001}
Ostrikov K, Yu M Y and Stenflo L 2001
\newblock On equilibrium states and dust charging in dusty plasmas
\newblock {\em IEEE Trans. Plasma Sci.} \textbf{29} 175--178

\bibitem{Land_Goedheer_2007}
Land V and Goedheer W J 2007
\newblock Manipulating dust charge using ultraviolet light in a complex plasma
\newblock {\em IEEE Trans. Plasma Sci.} \textbf{35} 280--285

\bibitem{Engwall_Eriksson_2006}
Engwall E, Eriksson A I and Forest J 2006
\newblock Wake formation behind positively charged spacecraft in flowing tenuous plasmas
\newblock {\em Phys. Plasmas} \textbf{13} 062904

\bibitem{Miloch_Vladimirov_2008}
Miloch W J, Vladimirov S V, P{\'e}cseli H L, and Trulsen J 2008
\newblock Wake behind dust grains in flowing plasmas with a directed photon  flux
\newblock {\em Phys. Rev.} E \textbf{77} 065401(R)

\bibitem{Miloch_Pecseli_Trulsen_2007}
Miloch W J, P{\'e}cseli H L and Trulsen J 2007
\newblock Numerical simulations of the charging of dust particles by contact  with hot plasmas
\newblock {\em Nonlin. Processes Geophys.} \textbf{14} 575--586

\bibitem{Miloch_Vladimirov_2008b}
Miloch W J, Vladimirov S V, P{\'e}cseli H L and Trulsen J 2008
\newblock Numerical simulations of potential distribution for elongated, insulating dust being charged by drifting plasmas
\newblock {\em Phys. Rev.} E \textbf{78} 036411

\bibitem{Lapenta_1999}
Lapenta G 1999
\newblock {Simulation of charging and shielding of dust particles in drifting plasmas}
\newblock {\em Phys. Plasmas} \textbf{6} 1442--47

\bibitem{Schott_1968}
Schott L 1968
\newblock Electrical probes.
\newblock In Lochte-Holtgreven W, editor, {\em Plasma Diagnostics}, pp.
  668--731 (Amsterdam: North Holland Publishing Company)

\bibitem{Tsytovich_Vladimirov_2004}
Tsytovich T and Vladimirov S 2004
\newblock Spinning of spherical grains in dusty plasmas
\newblock {\em IEEE Trans. Plasma Sci.} \textbf{32} 659--662

\bibitem{Piel_Melzer_2002}
Piel A and Melzer A 2002
\newblock Dynamical processes in complex plasmas
\newblock {\em Plasma Phys. Control. Fusion} \textbf{44} R1--R26

\bibitem{McDonald_Curry_2002}
McDonald K F, Curry R D and Hancock P J 2002
\newblock Comparison of pulsed and cw ultraviolet light sources to inactivate  bacterial spores on surfaces
\newblock {\em IEEE Trans. Plasma Sci.} \textbf{30} 1986--89

\bibitem{Laroussi_Mendis_2003}
Laroussi M, Mendis D A and Rosenberg M 2003
\newblock Plasma interaction with microbes
\newblock {\em New J. Phys.} \textbf{5} 41

\bibitem{Ewing_2000}
Ewing J J 2000
\newblock Excimer laser technology development
\newblock {\em IEEE J. Sel. Top. Quantum Electron.} \textbf{6} 1061

\bibitem{Neil_Meriminga_2002}
Neil G R and Merminga L 2002
\newblock Technical approaches for high-average-power free-electron lasers
\newblock {\em Rev. Mod. Phys.} \textbf{74} 685--701

\bibitem{Westley_Baragiola_1995}
Westley M S, Baragiola R A, Johnson R E and Baratta G A 1995
\newblock Photodesorption from low-temperature water ice in interstellar and circumsolar grains
\newblock {\em Nature} \textbf{373} 405--7

\bibitem{Baragiola_Vidal_2003}
Baragiola R A, Vidal R A, Svendsen W, Schou J, Shi M, Bahr D A and Atteberry C L 2003
\newblock Sputtering of water ice
\newblock {\em Nucl. Instr. Methods Phys. Res.} B \textbf{209} 294--303

\bibitem{Klumov_Morfill_2005}
Klumov B A, Morfill G E and Popel S I 2005
\newblock Formation of structers in dusty ionosphere
\newblock {\em J. Exp. and Theoret. Physics} \textbf{100} 152--164

\bibitem{Vishnakov_Trukhin_1991}
Vishnakov  V and Trukhin A 1991
\newblock Photoelectric emission and electronic structure of silicates:  SiO$_{2}$ and Na$_{2}$O $\cdot$ 3SiO$_{2}$.
\newblock {\em Nucl. Instr. Methods Phys. Res.} A  \textbf{308} 227--9

\end{thebibliography}

\end{document}